\documentclass[twocolumn]{aastex631}

\usepackage{csvsimple}
\usepackage{amsmath}
\usepackage{booktabs}
\usepackage[flushleft]{threeparttable}
\usepackage{hyperref}
\usepackage{xcolor}
\usepackage{comment}
\usepackage{enumitem}
\setcitestyle{authoryear,open={(},close={)}}

\usepackage{tikz} 

\def\HI{\rm H\,\textsc{i}}
\def\HIspace{\rm H\,\textsc{i} }

\def\kms{$\rm km~s^{-1}$}

\graphicspath{{./}{figures/}}

\begin{document}

\title{It's a Breeze: The Circumgalactic Medium of a Dwarf Galaxy is Easy to Strip}

\correspondingauthor{Jingyao Zhu}
\email{jingyao.zhu@columbia.edu}

\author[0000-0002-9001-6713]{Jingyao Zhu}
\affiliation{Department
of Astronomy, Columbia University, New York, NY 10027, USA}

\author[0000-0002-8710-9206]{Stephanie Tonnesen}
\affiliation{Center for Computational Astrophysics, Flatiron Institute, New York, NY 10010, USA}

\author[0000-0003-2630-9228]{Greg L. Bryan}
\affiliation{Department
of Astronomy, Columbia University, New York, NY 10027, USA}
\affiliation{Center for Computational Astrophysics, Flatiron Institute, New York, NY 10010, USA}

\author[0000-0002-1129-1873]{Mary E. Putman}
\affiliation{Department
of Astronomy, Columbia University, New York, NY 10027, USA}

\begin{abstract}
The circumgalactic medium (CGM) of star-forming dwarf galaxies plays a key role in regulating the galactic baryonic cycle. We investigate how susceptible the CGM of dwarf satellite galaxies is to ram pressure stripping (RPS) in Milky Way-like environments. In a suite of hydrodynamical wind tunnel simulations, we model an intermediate-mass dwarf satellite galaxy ($M_{*} = 10^{7.2}~M_{\odot}$) with a multiphase interstellar medium (ISM; $M_{\rm ISM} = 10^{7.9}~M_{\odot}$) and CGM ($M_{\rm CGM,vir} = 10^{8.5}~M_{\odot}$) along two first-infall orbits to more than 500 Myr past pericenter of a Milky Way-like host. The spatial resolution is $\sim$79 pc in the star-forming ISM and $316-632$ pc in the CGM. Our simulations show that the dwarf satellite CGM removal is fast and effective: more than $95\%$ of the CGM mass is ram-pressure-stripped within a few hundred Myrs, even under a weak ram pressure orbit where the ISM stripping is negligible. The conditions for CGM survival are consistent with the analytical halo gas stripping predictions in \cite{mccarthy_ram_2008}. We also find that including the satellite CGM does not effectively shield its galaxy, and therefore the ISM stripping rate is unaffected. Our results imply that a dwarf galaxy CGM is unlikely to be detected in satellite galaxies; and that the star formation of gaseous dwarf satellites is likely devoid of replenishment from a CGM.

\end{abstract}


\section{Introduction}\label{sec:intro}

Baryons fill the dark matter halo of a galaxy in the form of diffuse gas known as the circumgalactic medium (CGM). The CGM plays a key role in galaxy evolution, as it harbors a significant fraction of the baryonic mass and metals from the accretion of new gas and the expelled feedback material (see reviews by \citealt{putman_gaseous_2012,tumlinson_circumgalactic_2017,faucher-giguere_key_2023}). For the smallest galaxies in the Universe (dwarf galaxies; $\log M_{*} \lesssim 10^{9}~M_{\odot}$), the CGM has recently begun to be characterized in observations (\citealt{bordoloi_cos-dwarfs_2014,burchett_deep_2016,johnson_extent_2017}; see \citealt{zheng_comprehensive_2024} for a compilation). Though many properties remain to be determined, it is clear that many dwarf galaxies that are not near other galaxies have an extended CGM.

Dwarf galaxies that are satellites are extensively shaped by interactions with their hosts \citep{sales_baryonic_2022}, and their CGM will be the first baryons affected. The removal of satellite gas by the CGM of the host galaxy is known as ram pressure stripping (RPS; \citealt{gunn_infall_1972}). Cosmological simulations of Milky Way (MW) analogs expect that RPS, accompanied by gravitational effects \citep{mayer_tidal_2001,mayer_simultaneous_2006,mateo_velocity_2008,boselli_cold_2014,serra_meerkat_2023}, can remove the interstellar medium (ISM) and quench star formation for most dwarf satellites with $M_{*} \lesssim 10^{7}~M_{\odot}$ \citep{fillingham_taking_2015,simpson_quenching_2018,akins_quenching_2021,engler_satellites_2023,samuel_jolt_2023}. Observed dwarf satellites of spiral galaxies often show a lack of gas and star formation \citep{grcevich_h_2009,spekkens_dearth_2014,2015ApJ...808L..27W,carlsten_exploration_2022,zhu_census_2023} compared with field dwarfs at similar masses \citep{geha_stellar_2012,phillips_dichotomy_2014,karachentsev_morphological_2018}, although a large scatter has been found in the satellite populations and quenched fractions \citep{geha_saga_2017,bennet_m101_2019,mao_saga_2021,smercina_relating_2022}.

Satellite ISM removal by RPS has been widely observed in spiral galaxies in high-density environments (e.g., \citealt{van_gorkom_interaction_2004,sun_h_2007,ebeling_jellyfish_2014,kenney_transformation_2014,poggianti_gasp_2017,jachym_alma_2019}; see recent reviews by \citealt{cortese_dawes_2021,boselli_ram_2022}). The ISM stripping efficiency and the impact on satellite star formation have been analyzed in numerical simulations (e.g., \citealt{abadi_ram_1999,schulz_multi_2001,roediger_ram_2005,kapferer_effect_2009,tonnesen_gas_2009,tonnesen_star_2012,bahe_competition_2012,bekki_galactic_2014,lee_dual_2020,troncoso-iribarren_better_2020,rohr_jellyfish_2023,zhu_when_2024}) as well as analytical models \citep{fujita_effects_1999,koppen_ram_2018}. Dwarf satellite galaxies, on the other hand, are more likely already gasless and quenched 
because of their shallow gravitational potential. Direct observations of dwarf stripping are limited and require targeted deep observations (e.g., \citealt{pearson_local_2016,kleiner_meerkat_2023}). Idealized, controlled simulations have investigated dwarf satellite RPS in relatively low-density environments, e.g., the outskirts of groups and clusters \citep{mori_gas_2000,steyrleithner_effect_2020} and MW-like environments \citep{mayer_simultaneous_2006,gatto_unveiling_2013,salem_ram_2015,emerick_gas_2016}. However, the simulations have found a mixture of dwarf ISM stripping efficiencies. Including physical processes such as gas outflows due to star formation is shown to be essential for realistic RPS outcomes \citep{emerick_gas_2016,garling_dual_2024}. 

Dwarf galaxy stripping simulations have largely not included a CGM in the models. These simulations, therefore, lack information on the fate of satellite CGM: is the CGM ram-pressure-stripped before ISM removal begins? If the CGM stripping is effective, it will cause a lack of future gas accretion, affect the satellite's star formation, and contribute significant amounts of gas to the host halo. Previous simulations on spiral or elliptical galaxy halo gas stripping found that the CGM removal is often incomplete \citep{mccarthy_ram_2008,bekki_ram-pressure_2009,roediger_stripped_2015,vijayaraghavan_ram_2015}, which may explain the X-ray bright tails observed in some galaxies in groups and clusters \citep{sun_x-ray_2007,jeltema_hot_2008,hou_x-ray_2024}. In the dwarf galaxy regime, there is a lack of CGM stripping studies except for the Large Magellanic Cloud (LMC): the observed ionized part of the Magellanic Stream \citep{fox_kinematics_2020} could consist of the stripped remnant of the LMC's CGM \citep{lucchini_magellanic_2020,lucchini_magellanic_2021,krishnarao_observations_2022}; but see \cite{kim_identifying_2024}. However, dwarf satellites as massive as the LMC ($M_{*} = 2.7 \times 10^{9}~M_{\odot}$; \citealt{vandermarel_new_2002}) are relatively uncommon in a cosmological context \citep{liu_how_2011,tollerud_small-scale_2011,robotham_galaxy_2012,engler_abundance_2021}. There is a need for constraining the CGM stripping efficiency for intermediate-mass dwarf galaxies ($M_{*} \approx M_{\HI} \in 10^{7}-10^{7.5}~M_{\odot}$) that are common among observed gaseous/star-forming dwarf satellites and the quenched fractions the most unconstrained 
($20-100\%$; \citealt{geha_saga_2017,bennet_m101_2019,mao_saga_2021,carlsten_exploration_2022,karunakaran_h_2022,zhu_census_2023}).

In this work, we investigate the RPS of an intermediate-mass dwarf satellite galaxy with an extended CGM. In a suite of hydrodynamical \textit{wind tunnel} simulations, we vary the satellite ram pressure as in a MW-like host's first-infall orbits until more than 500 Myr past pericenter, and we vary whether the satellite CGM is included. The simulation results constrain the timescale and efficiency of the dwarf galaxy ISM \textit{and} CGM removal. Comparison between the cases with and without a satellite CGM additionally quantifies the effect of the satellite CGM on the ISM stripping rate. In the simulations, we model radiative cooling in the multiphase gas, star formation, and supernovae feedback to account for the outflows, independent of RPS. We compare our results with analytical predictions of dwarf galaxy and halo gas RPS and discuss the implications for dwarf galaxy observations. 

The paper is organized as follows. Section \ref{sec:methods} outlines our methodology: the dwarf galaxy model (\S \ref{subsec:dwarf}), the orbits and ram pressure profiles (\S \ref{subsec:orbits}), and the simulation suite (\S \ref{subsec:sims}). Section \ref{analytical_solutions} reviews two analytical predictions in the literature, the dwarf galaxy central gas stripping criterion (\S \ref{subsec:analytic_core_strip}; \citealt{mori_gas_2000}) and the CGM stripping criterion (\S \ref{subsec:analytic_cgm_strip}; \citealt{mccarthy_ram_2008}). Section \ref{sec:results} covers the main results of our simulations. \S \ref{subsec:gas_removal} describes the RPS morphology, timescale, and efficiency, \S \ref{subsec:cgm_shielding} addresses whether the inclusion of satellite CGM affects the ISM stripping rate, and \S \ref{subsec:stripping_criteria} presents the gas surface density radial profiles under RPS and compares our simulations with the analytical criteria (\S \ref{analytical_solutions}). Section~\ref{sec:discussion} discusses our results in a broader context, and Section~\ref{sec:summary} summarizes the main findings.

\section{Methodology} \label{sec:methods}

We run a suite of three-dimensional dwarf galaxy \textit{wind tunnel} simulations using the adaptive mesh refinement (AMR) code Enzo \citep{bryan_enzo_2014}. The dwarf galaxy is placed in a $162^{3}$ kpc simulation volume with a $256^{3}$ root grid resolution. We allow up to three levels of refinement so that the highest spatial resolution is $79$ pc. In the cases where a dwarf satellite CGM is included (\S \ref{subsec:dwarf}), the typical resolution within the CGM is 316--632 pc, or at least 190--380 cells across the CGM diameter. We model the radiative cooling of the multiphase gas using the Grackle chemistry and cooling library\footnote{\url{https://grackle.readthedocs.io/}} \citep{smith_grackle_2017}, which calculates photoheating and photoionization from the UV background of \cite{haardt_radiative_2012}. We use the star formation recipe of \cite{goldbaum_mass_2015} and the stellar and supernovae feedback model from \cite{goldbaum_mass_2016}; also see our previous work for details \citep{zhu_when_2024}.

\subsection{The dwarf satellite galaxy: initial conditions}\label{subsec:dwarf}
This section outlines the mass and structural properties of our modeled dwarf satellite galaxy. We select the stellar mass to be $M_{*} = 10^{7.2}~M_{\odot}$, which is within the most common mass range of observed star-forming dwarf satellites and where the quenched fraction is the most uncertain (\citealt{geha_saga_2017,bennet_m101_2019,mao_saga_2021,carlsten_exploration_2022,karunakaran_h_2022,zhu_census_2023}; see the Introduction). Because dwarf galaxies' dark matter mass and distribution are generally uncertain \citep{oman_unexpected_2015,sales_baryonic_2022}, we base the satellite galaxy's initial conditions on the Local Group dwarf irregular Wolf–Lundmark–Melotte (WLM), which has excellent literature data (e.g., \citealt{leaman_resolved_2012,oh_high-resolution_2015}). Besides the initial mass and structure, we do not attempt to model WLM as it is relatively isolated and not an infalling satellite \citep{mcconnachie_solo_2021,putman_gas_2021}. The dwarf galaxy parameters are summarized in Table \ref{table:dwarf_model} and detailed below.

\vspace{-2.5em}
\begin{deluxetable*}{cccccccccccccccc}\label{table:dwarf_model}
\tablecaption{Initial mass and structural parameters of the dwarf satellite galaxy}
\tablehead{\multicolumn{3}{c}{Stellar Disk}  &  & \multicolumn{3}{c}{Dark Matter} & &  \multicolumn{3}{c}{Gas Disk} &  & \multicolumn{4}{c}{Gaseous Halo}               \\
\cline{1-3}\cline{5-7}\cline{9-11}\cline{13-16} 
$M_{*}$ & $a_{*}$ & $b_{*}$ &  & $M_{\rm vir}$ & $\rho_{d0}$ & $r_{0}$ &  & $M_{\rm gas}$ & $a_{\rm gas}$ & $b_{\rm gas}$ & &  $M_{\rm CGM,vir}$ & $\alpha$ & $r_{g0}$ & $\rho_{g0}$\\
($M_{\odot}$) &	(kpc)   & (kpc)   &  & ($M_{\odot}$) & ($\rm g~cm^{-3}$)  &  (kpc)	&  & ($M_{\odot}$) & (kpc)	       &	(kpc) &   &   ($M_{\odot}$)  &  & (kpc) & ($\rm g~cm^{-3}$) }
\startdata
$10^{7.2}$	  &	0.75	&	0.375 &  & 	$10^{9.9}$ &  1.82e-24	      &	2.2     &  & $10^{7.9}$	   &	1.04	   & 0.52  &   & $10^{8.5}$    &	 $-1.5$ &  5.2  & 8.26e-28 
\enddata
\tablecomments{Each parameter is described in Section \ref{subsec:dwarf}; also see equation \ref{eqn:cgm_Menclosed} for the gaseous halo parameters.}
\end{deluxetable*}

We model the first three components in Table \ref{table:dwarf_model} as in our previous work \citep{zhu_when_2024}: the stellar disk and dark matter as static gravitational potential fields, and the gas disk evolved with an AMR grid. The static stellar potential follows the Plummer–Kuzmin model \citep{miyamoto_three-dimensional_1975} and the gas is initialized as a smoothed exponential disk, where the input masses ($M_{*}$, $M_{\rm gas}$) and scale radii ($a_{*}$, $a_{\rm gas}$) are from \cite{read_understanding_2016}. The stellar mass grows from the initial value as star formation proceeds in the simulations. The scale heights ($b_{*}$, $b_{\rm gas}$) are obtained under a height-to-radius ratio of $0.5$ from dwarf galaxy intrinsic shape measurements \citep{kado-fong_tracing_2020}. We adopt a spherical Burkert model \citep{burkert_structure_1995,mori_gas_2000} for the cold dark matter potential using the WLM maximum rotation velocity $v_{\rm max}=39$ \kms \citep{read_understanding_2016}. The resulting dark matter core density ($\rho_{d0}$) and core radius ($r_{0}$) are listed in Table \ref{table:dwarf_model}, and the virial radius and enclosed dark matter mass ($r_{\rm vir} \approx 42$ kpc, $M_{\rm vir} \approx 10^{9.9}$ $M_{\odot}$) are consistent with those under a \textsc{core}NFW model \citep{read_understanding_2016}.

For the dwarf gaseous halo (CGM), we first model an isothermal sphere with a power-law density profile, $\rho(r) = \rho_{g0} (r/r_{g0})^{\alpha}$, where $\rho_{g0}$ is the CGM scaling density at a scaling radius $r_{g0}$, and $\alpha$ is the power-law index (Table \ref{subsec:dwarf}). The enclosed CGM mass is given by, 
\begin{equation}\label{eqn:cgm_Menclosed}
    M_{\rm CGM,enc}(r) = \int_{r_{\rm min}}^{r} 4 \pi r^{2} \rho_{g0} \cdot (r/r_{g0})^{\alpha} dr
\end{equation}
where $r_{\rm min}$ is the CGM inner boundary radius. We select the power-law index $\alpha=-1.5$ based on \cite{stern_cooling_2019} and $r_{\rm min}=5 a_{\rm gas} = 5.2$ kpc to be the ISM-CGM transition radius. Adopting the CGM scaling density and radius in Table \ref{table:dwarf_model}, the enclosed CGM mass within $r_{\rm vir}$ is $M_{\rm CGM,enc}(r_{\rm vir}) = 10^{8.5}~M_{\odot}$. This mass corresponds to $25\%$ of the baryonic mass budget within $r_{\rm vir}$, which is around the upper limit found in the $M_{\rm vir} \approx 10^{10}~M_{\odot}$ dwarf galaxies in cosmological simulations \citep{christensen_-n-out_2016,hafen_origins_2019}, and much higher than the cold phase CGM mass in local volume observations \citep{zheng_comprehensive_2024}. We choose to model this relatively massive satellite CGM for the purpose of testing the CGM's susceptibility to RPS --- a less massive CGM may be a better match to 
isolated dwarf galaxies but will be more easily removed under stripping.

\begin{figure*}[!htb]
    \centering
    \includegraphics[width=1.0\linewidth]{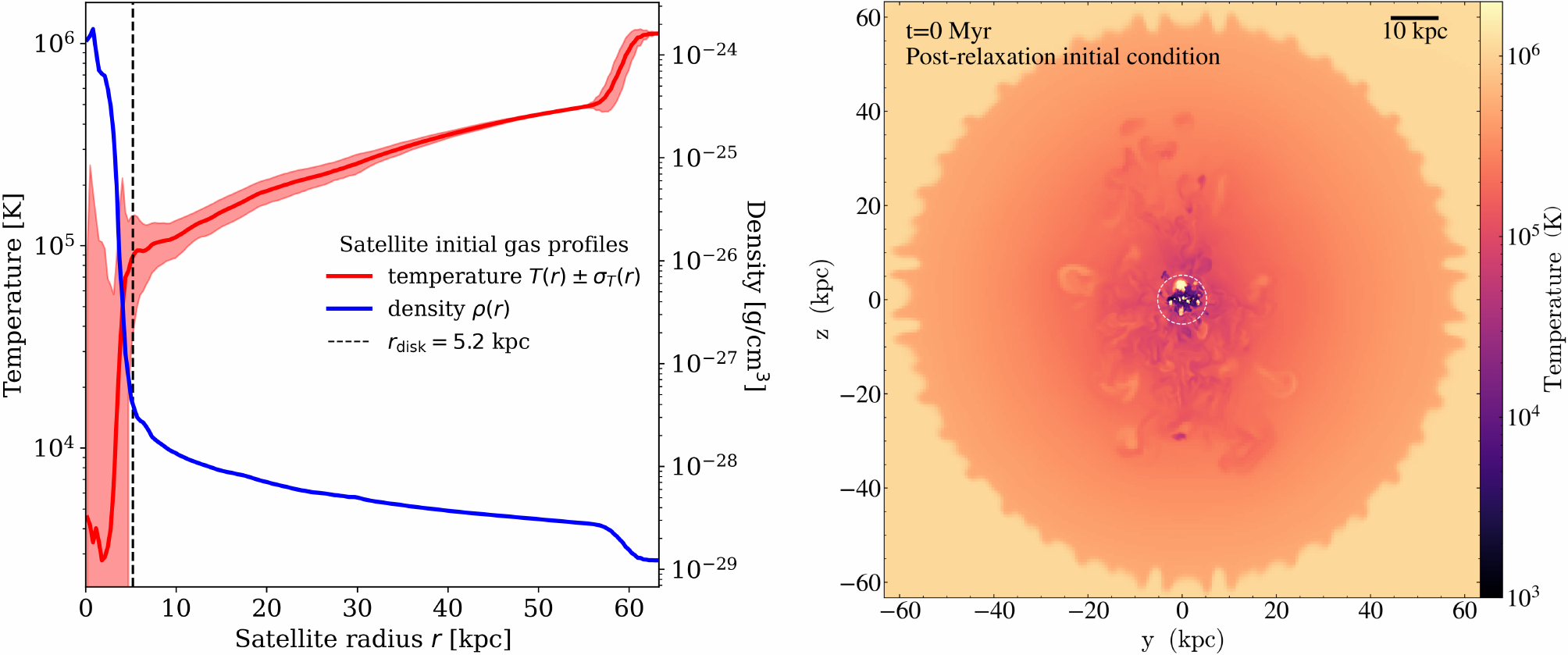}
    \caption{Simulation initial conditions. Left: temperature and density radial profiles at $t=0$ Myr, defined as the moment wind enters the box after the CGM relaxation phase (\S \ref{subsec:dwarf}). The solid lines show the density and temperature averages on the right- and left-hand y-axis, respectively, and the red-shaded region shows the temperature standard deviation ($\pm \sigma_{T}$). The dwarf CGM extends beyond $r_{\rm vir} \approx 42$ kpc to about 60 kpc. Right: gas temperature slice (at $x=0$ kpc, edge-on). The dashed white circle annotates the disk-halo transition radius ($r_{\rm disk}=5.2$ kpc; see dashed line in the left panel and \S \ref{subsec:dwarf}). Gas in the disk and inner halo is turbulent and multiphase.}
    \label{fig:dwarf_init_condition}
\end{figure*}

Once the simulations begin, the dwarf galaxy model in Table \ref{table:dwarf_model} is subject to an initial relaxation phase of 2 Gyr, allowing the galaxy to gradually radiatively cool. The rotating gas disk collapses and forms stars, driving stellar and supernova feedback, and ejecting low-density, metal-enriched\footnote{Due to the lack of strong observational constraints, we set the initial gas metallicity to $Z_{\rm gas}=0.3 Z_{\odot}$ within the disk ($r<r_{\rm disk}=5.2$ kpc) and $0.1 Z_{\odot}$ in the halo. Metals produced by star formation can be mixed into the surrounding gas via feedback.} material into the halo. The initially static, isothermal, and smooth CGM also cools and develops a temperature gradient; after interactions with outflows from the gas disk, it eventually becomes turbulent and multiphase. The post-relaxation dwarf galaxy with a multiphase CGM is shown in Figure \ref{fig:dwarf_init_condition}; it is used as the initial condition for the wind tunnel simulations (\S \ref{subsec:sims}).

The left panel of Figure \ref{fig:dwarf_init_condition} shows the satellite dwarf's post-relaxation density and temperature profiles used as initial conditions. The density profile $\rho(r)$ (blue solid line and right-hand y-axis) decreases steeply within the disk region ($r \leq r_{\rm disk}$; vertical dashed line) and flattens in the halo region to a post-relaxation power-law index $\alpha' \approx -1$ (Table \ref{table:dwarf_model}). Our modeled dwarf CGM extends beyond $r_{\rm vir}$ to $r \approx 60$ kpc, resulting in a total enclosed mass $M_{\rm CGM,enc}(r \approx 60~\rm kpc) \approx 10^{8.67}~M_{\odot}$ (equation \ref{eqn:cgm_Menclosed}). In later sections, we consistently adopt $1.5 r_{\rm vir} = 63$ kpc as the CGM upper radius limit to account for its total mass (e.g., Figure \ref{fig:global_time_evolution}).

Figure \ref{fig:dwarf_init_condition}'s right panel shows the temperature structure, which is quantified by the radial profile with scatter in the left panel ($T(r) \pm \sigma_{T}(r)$; red-shadowed line and left-hand y-axis). Gas is multiphase within $r \lesssim 30$ kpc, as seen from the slice and the large scatter in the radial profile; it consists of the cold ISM and hot feedback bubbles within the disk (white dashed circle) and the inner CGM that can be heated by feedback or cool and inflow into the ISM. The multiphase gas ($r \lesssim 30$ kpc) is also turbulent. Our modeled CGM is not in thermal or static equilibrium, reflected by a continuous net inflow into the ISM in the absence of external interactions like RPS (see Figure \ref{fig:global_time_evolution}; will be discussed in \S \ref{sec:results}). We address these systematics by modeling isolated control cases in addition to RPS cases (Table \ref{tab:simulation_suite}).

Finally, dwarf galaxies are not cold rotating disks, as turbulent gas velocities are often more comparable with the rotational velocities than in larger disk systems (e.g., \citealt{wheeler_no-spin_2017,lelli_gas_2022}). 
We model an initially rotation-supported gas disk, and the stellar and supernova feedback adds to the dispersion support once star formation occurs. Our choice of a relatively low volume density threshold for star formation, $n_{\rm SF} = 1$ cm$^{-3}$, was made to allow for radially extended star formation across the $\sim$5 kpc disk so that feedback creates realistic gas velocity dispersions in the ISM (the \HI-weighted dispersion is $\sigma_{\HI} \approx 7.3$ km/s at the $t=0$ snapshot in Figure \ref{fig:dwarf_init_condition}). This value and the resulting star formation rate (SFR) in our modeled galaxy of $\rm SFR_{t=0} \approx 2.5 \times 10^{-3}~M_{\odot}/\rm yr$ is consistent with observations of WLM-mass dwarf galaxies \citep{leroy_star_2008,zhang_outside-shrinking_2012,mcgaugh_star-forming_2017}.

\subsection{Satellite orbits and ram pressure profiles}\label{subsec:orbits}
We model realistic, time-dependent ram pressure profiles of infalling dwarf satellites in MW-like spiral host halos. We sample two cases, (i) a fiducial case representing the most likely first infall orbits; and (ii) a stripping threshold case derived from analytical predictions (\S \ref{analytical_solutions}), where the peak ram pressure is predicted to remove the dwarf CGM described in \S \ref{subsec:dwarf}.

Since ram pressure is defined as $P_{\rm ram} = \rho_{\rm host} \cdot v_{\rm sat}^{2}$ \citep{gunn_infall_1972}, varying ram pressure can be achieved by varying the density of the host's stripping medium ($\rho_{\rm host}$, here the spiral host galaxy's CGM) or the satellite orbital velocity ($v_{\rm sat}$). In this work, we vary satellite orbits while fixing the spiral CGM model to that in our previous work \citep{zhu_when_2024}. 
The host stripping medium ($\rho_{\rm host}$) follows a modified MW CGM density profile under the \cite{miller_constraining_2015} parametrization and is boosted by a constant factor of $C=2.73$ to address the potential density underestimation, particularly at low orbital radii (\citealt{salem_ram_2015,voit_ambient_2019}).

We numerically integrate the satellite orbits using the Galactic Dynamics package Gala \citep{price-whelan_gala_2017,price-whelan_adrngala_2020}, where the dwarf satellite is approximated as a point mass traveling in a spiral host with a Navarro–Frenk–White (NFW) dark matter potential \citep{navarro_structure_1996}. We adopt the host halo mass $M_{200,\rm host}=1.5 \times 10^{12}~M_{\odot}$ based on literature values of the MW \citep{sawala_local_2023} and NFW concentration $c=10$ \citep{ludlow_mass-concentration-redshift_2014}. The resulting virial radius of the spiral host is $R_{200,\rm host} = 242$ kpc.

\vspace{-2.5em}
\begin{deluxetable*}{cccccccc}\label{tab:orbits}
\tablecaption{Orbital parameters of the dwarf satellite in a Milky Way-like spiral halo} 
\tablehead{\colhead{Orbit} & \colhead{$e$} & \colhead{($|V_{\phi,0}|, |V_{r,0}|)$} & \colhead{$R_{\rm peri}$} & \colhead{$n_{\rm peri}$} & \colhead{$v_{\rm peri}$} & \colhead{$P_{\rm ram,peak}$}  &   \colhead{$\tau_{\rm total}$}\\
\colhead{} & \colhead{} & \colhead{($V_{200}$)} & \colhead{(kpc)} & \colhead{(cm$^{-3}$)} & \colhead{(\kms)} & \colhead{(dyne/cm$^{2}$)} & \colhead{(Myr)}\\
\colhead{(1)} & \colhead{(2)} & \colhead{(3)} & \colhead{(4)} & \colhead{(5)} & \colhead{(6)} & \colhead{(7)} & \colhead{(8)}}
\startdata
fiducial	&	0.85	&	(0.402, 1.037)	&	40	&	1.49e-4	&	399	&	3.94e-13	&	1900	\\
weak	    &	0.6	    &	(0.8, 0.772)	&	110	&	3.22e-5	&	286	&	4.38e-14	&	2296	\\
\enddata
\tablecomments{Columns 1--2: the two orbits simulated in this work and the eccentricities. Column 3: the initial tangential ($|V_{\phi,0}|$) and radial ($|V_{r,0}|$) velocity magnitudes at $R_{200}$, both in units of the host's virial velocity ($V_{200} \equiv \sqrt{G \cdot M_{200}/R_{200}} = 163$ \kms). Columns 4--7: the pericentric distance ($R_{\rm peri}$), host stripping medium number density ($n_{\rm peri}$), satellite velocity ($v_{\rm peri}$), and the peak ram pressure strength ($P_{\rm ram,peak}= m_{u} n_{\rm peri} \cdot v_{\rm peri}^{2}$). Column 8: the satellite's total traveling time from the host's $R_{200}$ to $R_{\rm peri}$ (first infall segment) and back to $R_{200}$ (post-pericenter segment).}
\end{deluxetable*}

Table \ref{tab:orbits} summarizes the key parameters of the two orbits, one representing the most probable infalling dwarf satellite (\textit{fiducial}), and the other the minimum $P_{\rm ram}$ orbit predicted to strip the dwarf CGM (\textit{weak}, because of its low peak $P_{\rm ram}$; see \S \ref{analytical_solutions} for details). The fiducial orbit adopts the most probable eccentricity $e=0.85$ for infalling satellites in cosmological simulations \citep{wetzel_orbits_2011}, resulting in a pericentric distance of $R_{\rm peri}=40$ kpc, which agrees well with observed MW infalling dwarf satellites \citep{fritz_gaia_2018,putman_gas_2021}. The weak $P_{\rm ram}$ orbit in Table \ref{tab:orbits} shares the same initial velocity magnitude as the fiducial run but at a lower eccentricity $e=0.6$ (lower radial velocity component $|V_{r}|$; Table \ref{tab:orbits}) such that its $P_{\rm ram,peak} \approx 4 \times 10^{-14}$ dyne/cm$^{2}$ is the predicted minimum value for CGM removal (\S \ref{analytical_solutions}). Its pericentric distance of $R_{\rm peri}=110$ kpc is greater than most MW satellites \citep{putman_gas_2021}, implying that the $P_{\rm ram}$ threshold for CGM removal is lower than the peak $P_{\rm ram}$ of typical infalling dwarf satellites (and similarly, less eccentric than most satellites in cosmological simulations; \citealt{wetzel_orbits_2011}).

The orbits consist of a first-infall segment (host $R_{200}$ to $R_{\rm peri}$) followed by a post-pericenter segment ($R_{\rm peri}$ to $R_{200}$; see, e.g., \citealt{simpson_quenching_2018,fillingham_environmental_2018}). We include the post-pericenter segment because of the previous finding that mass transport can be sensitive to the orbital history, i.e., the time derivative of ram pressure \citep{tonnesen_journey_2019,zhu_when_2024}. Ram pressure peaks at the orbital pericenter, where the host CGM densities and satellite velocities are listed in Table \ref{tab:orbits}. The orbital ram pressure time evolution read in from simulation outputs ($\rho_{\rm gas} \cdot v_{\rm gas}^2$ in the wind direction) is shown in Figure \ref{fig:global_time_evolution}.

\subsection{The simulation suite}\label{subsec:sims}

In the simulation suite, we vary whether the dwarf satellite CGM (\S \ref{subsec:dwarf}) is included; when the CGM is included, we additionally vary the ram pressure strength (\S \ref{subsec:orbits}) to test the CGM removal conditions. Table \ref{tab:simulation_suite} summarizes each simulation's short name and setup.

\vspace{-2.5em}
\begin{deluxetable*}{cccc}\label{tab:simulation_suite}
\tablecaption{Overview of the simulation suite} 
\decimalcolnumbers
\tablehead{\colhead{Simulation} & \colhead{Wind} & \colhead{Sat. CGM} & \colhead{Comment}}
\startdata
\texttt{MW-w}	    &	fiducial	&	Y        & Satellite with CGM under the MW fiducial wind.       \\
\texttt{weak-w}	    &	weak	    &	Y        & Satellite with CGM under the low ram pressure wind.      \\
\texttt{iso}	    &	\nodata	    &	Y        & Satellite with CGM in isolation (no wind).           \\
\texttt{MW-w nocgm}	&	fiducial	&	N        & Satellite without CGM (ISM only) under the MW fiducial wind.   \\
\texttt{iso nocgm}  &   \nodata     &   N        & Satellite without CGM in isolation. \\
\enddata
\tablecomments{Column 1 lists the simulation short names used throughout this paper. Column 2 shows the wind cases, where the fiducial and weak wind orbits (\S \ref{subsec:orbits}, Table \ref{tab:orbits}). Column 3 states whether (Y/N) the case includes the satellite CGM (\S \ref{subsec:dwarf}, Table \ref{table:dwarf_model}). Column 4 summarizes the simulation setup.}
\end{deluxetable*}

In the following sections, we will compare the simulations in the suite (Table \ref{tab:simulation_suite}) as follows. The two stripping cases with CGM (\texttt{MW-w}, \texttt{weak-w}) constrain the stripping rate and final fate of the satellite CGM (\S \ref{subsec:gas_removal}), offering a direct comparison with analytical stripping criteria (\S \ref{subsec:stripping_criteria}). The non-equilibrium state of the dwarf CGM leads to temporal trends (net inflow from inner halo to disk; CGM to ISM mass transfer; Sections \ref{subsec:dwarf} and \ref{sec:results}) even in the absence of RPS, which is quantified in \texttt{iso}. By comparing \texttt{iso} with the two stripping cases, we can separate the internal effects from RPS and make predictions for field versus satellite dwarfs (\S \ref{sec:discussion}). 

We also compare the ISM stripping of the cases with and without the satellite CGM. The two \texttt{nocgm} cases adopt the same initial stellar, dark matter, and gas disk properties in Table \ref{table:dwarf_model} but without the gaseous halo. The CGM inflows during the relaxation phase (\S \ref{subsec:dwarf}) result in a slightly higher initial ISM mass in the cases with CGM than their \texttt{nocgm} counterparts (Figure \ref{fig:global_time_evolution}). Comparing \texttt{MW-w} and \texttt{MW-w nocgm} quantifies the potential impact of the CGM on the ISM stripping rate (\S \ref{subsec:cgm_shielding}). 

We implemented two Eulerian fluid tracers to track gas motion and mixing: ISM and CGM colors, denoting the ISM and CGM density fractions, respectively, within each gas cell. Initially, the ISM (CGM) fraction is 1 (0) in the disk region and 0 (1) in the halo, whereas both fractions are 0 in the ambient gas (Figure \ref{fig:dwarf_init_condition} at $r \geq 60$ kpc) and the ram pressure wind. In \texttt{iso}, for example, the disk region CGM fraction gradually increases as the inner halo gas cools and inflows, and the halo region ISM fraction also increases because of feedback ejections. During RPS, the color tracers track the percentage of the survived/stripped gas originating from the disk or halo; see \S \ref{sec:results}, Figure \ref{fig:morphology} for detailed usage of the tracers.

\section{Analytical predictions: RPS and dwarf satellite gas removal}\label{analytical_solutions}
In this section, we apply analytical RPS criteria from the literature to the parameter space in this work --- dwarf galaxy satellite RPS in a spiral galaxy host (\S \ref{sec:methods}). Section \ref{subsec:analytic_core_strip} predicts the stripping of central dense gas \citep{mori_gas_2000}, i.e., whether RPS can remove all the gas from the dwarf satellite. Section \ref{subsec:analytic_cgm_strip} predicts the stripping of spherical diffuse gas \citep{mccarthy_ram_2008}, i.e., to what radius the satellite's CGM can be removed. The analytical predictions will be compared with our numerical results in the following sections (\S \ref{sec:results} and \ref{sec:discussion}).

\subsection{Central dense gas stripping}\label{subsec:analytic_core_strip}
Instantaneous RPS conditions are generally defined as where the ram pressure exceeds the gravitational restoring force in the satellite. For example, the \cite{gunn_infall_1972} criterion approximates spiral satellite galaxies as two-dimensional thin disks whose gravitational force is dominated by the stellar component perpendicular to the disk. Dwarf galaxies, however, are more spherical and dark matter dominated \citep{behroozi_universemachine_2019,kado-fong_tracing_2020,carlsten_structures_2021}. We follow the \cite{mori_gas_2000} analytical prescription, which assumes the dwarf's gravitational restoring force is dominated by its spherical dark matter core. The value of $P_{\rm ram}$ needed to exceed gravity in the core region, which equals the thermal pressure $P_{\rm thermal,0} = \rho_{\rm gas,0} k_{B} T/(\mu m_{p})$ under the simplified assumption of hydrostatic equilibrium, is given by,

\begin{equation}\label{eqn:burkert_pressure_balance}
    P_{\rm ram} = \rho_{\rm host} \cdot v_{\rm sat}^{2} > \frac{\rho_{\rm gas,0} k_{B} T}{\mu m_{\rm u}} = \frac{G M_{0} \rho_{\rm gas,0}}{3 r_{0}}
\end{equation}
where $M_{0}$, $r_{0}$ is the mass and radius of the dark matter core and $\rho_{\rm gas,0}$ is the core gas density (see \citealt{mori_gas_2000}). If we further assume that the dwarf galaxy's dark matter halo follows the one-parameter Burkert model \citep{burkert_structure_1995}, as used in this work (\S \ref{subsec:dwarf}), the core radius and gas density both scale with the core dark matter mass, $r_{0} \propto M_{0}^{3/7}$ and $\rho_{\rm gas,0} \propto \rho_{d0} \propto M_{0}^{-2/7}$. The pressure balance criterion (equation \ref{eqn:burkert_pressure_balance}) can be rewritten as a core dark matter mass ($M_{0}$) threshold,

\begin{equation}\label{eqn:burkert_core_criterion}
    \frac{P_{\rm ram}}{1.66 \times 10^{-12} \rm ~dyne/cm^{2}} > \left(\frac{F}{0.1}\right) \left(\frac{M_{\rm 0,crit}}{1.27 \times 10^{9} ~M_{\odot}}\right)^{2/7}
\end{equation}

where $F = M_{\rm gas,0}/M_{0}$ is the core gas mass divided by the core dark matter mass, and $M_{\rm 0,crit}$ the critical core dark matter mass below which the central gas can be removed under $P_{\rm ram}$ (see \citealt{mori_gas_2000} equation 13). The ratio $F$ remains close to the expected value $0.1$ in our simulations, and given the core mass of the WLM-mass dwarf satellite ($M_{0}=4.6 \times 10^{8}~M_{\odot}$; obtained from the value of $v_{\rm max}$ in \S \ref{subsec:dwarf}), equation \ref{eqn:burkert_core_criterion} predicts that $P_{\rm ram} > 1.24 \times 10^{-12}$ dyne/cm$^{2}$ is required for central gas removal. This $P_{\rm ram}$ is three times higher than the peak value in our simulations ($P_{\rm ram,peak}$ of \texttt{MW-w}; Table \ref{tab:orbits}), which implies that the central dense gas in our WLM-mass satellite cannot be removed in the MW fiducial orbit based on \cite{mori_gas_2000}.

In this work, we selected to model a cored dark matter profile under the \cite{burkert_structure_1995} parameterization (\S \ref{subsec:dwarf}) because it matches observational rotation curves for dwarf galaxies \citep{de_blok_high-resolution_2008,oh_high-resolution_2015}. However, dwarf galaxy rotation curves show a large diversity \citep{oman_unexpected_2015,sales_baryonic_2022}, and the rotation curves can be successfully reproduced under other dark matter models (e.g., \citealt{di_cintio_mass-dependent_2014,read_understanding_2016}). Different choices of dark matter models will modify the stripping criterion (equation \ref{eqn:burkert_pressure_balance}) from the special form in equation \ref{eqn:burkert_core_criterion}. For example, a ``cuspy" NFW model adopts a steeper gravitational restoring force profile in the galaxy center, hence requires higher $P_{\rm ram}$ for the central gas removal than the cored Burkert model, but the difference is shown to be negligible \citep{emerick_gas_2016}.

\subsection{Spherical diffuse gas stripping}\label{subsec:analytic_cgm_strip}
We now consider the instantaneous RPS criterion for the dwarf galaxy CGM, which is low in density, but high in mass and volume compared with the central dense gas (\S \ref{subsec:dwarf}). We follow the analytical prescription of \cite{mccarthy_ram_2008}, originally for galaxy CGM stripping in groups and clusters, and apply it to the dwarf satellite in spiral host parameter space. The \cite{mccarthy_ram_2008} criterion, similar to the central gas RPS criterion above (equation \ref{eqn:burkert_pressure_balance}), compares ram pressure with the gravitational restoring force, but here for a spatially extensive distribution. The value of $P_{\rm ram}$ required to balance the gravitational restoring force per unit area ($F_{\rm grav}/dA$) as a function of the projected distance to the galaxy center ($R$; projected along the ram pressure direction) is given by,

\begin{equation}\label{eqn:mccarthy_pressure_balance}
\begin{split}
    P_{\rm ram} > F_{\rm grav}/dA & = a_{\rm grav,max}(R) \cdot \Sigma_{\rm gas}(R)\\
    & = \alpha \frac{G M_{\rm tot}(R) \cdot \rho_{\rm gas}(R)}{R}
\end{split}
\end{equation}
where $a_{\rm grav,max}$ is the maximum gravitational restoring acceleration along the projection, and $M_{\rm tot}$, $\Sigma_{\rm gas}$, and $\rho_{\rm gas}$ are the total mass (dominated by dark matter), gas projected density, and gas volume density, respectively \citep{mccarthy_ram_2008}. For a spherically symmetric gravitational potential, the restoring acceleration $a_{\rm grav,max}$ is usually maximized where gravity is 45$^{\circ}$ anti-aligned with ram pressure. The $\alpha$ factor in equation~\ref{eqn:mccarthy_pressure_balance} is a geometric factor of order unity from integration along the projection \citep{mccarthy_ram_2008}.

\begin{figure}[!htb]
    \centering
    \includegraphics[width=1.0\linewidth]{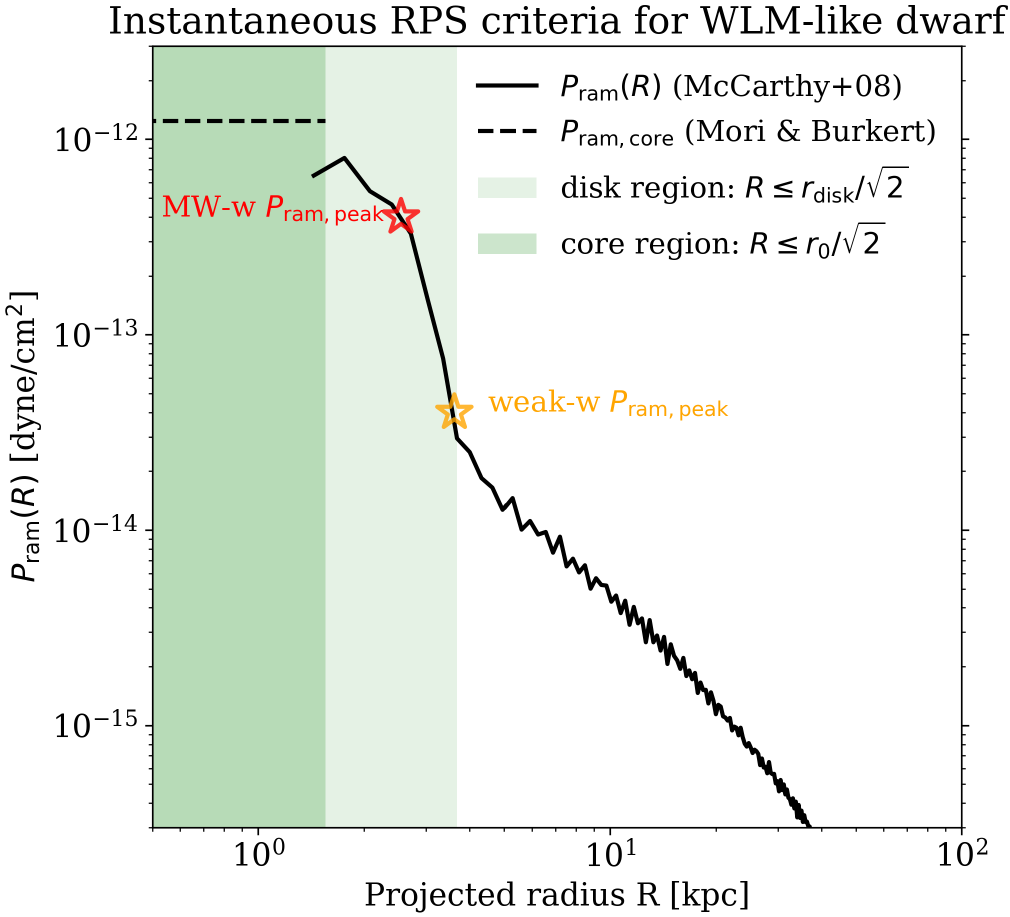}
    \caption{Ram pressure thresholds ($P_{\rm ram,thresh}$) predicted for the WLM-like dwarf galaxy stripping (\S \ref{analytical_solutions}). The solid curve shows the radial profile of $P_{\rm ram,thresh}$ across the entire dwarf galaxy (\citealt{mccarthy_ram_2008}; see equation \ref{eqn:mccarthy_pram_threshold}), and the dashed horizontal line shows the core stripping criterion (\citealt{mori_gas_2000}; see equation \ref{eqn:burkert_core_criterion}). The lighter- and darker-shaded regions show the projected disk and core regions, respectively. Where the solid curve intersects with $R=r_{\rm disk}/\sqrt{2}$ is the predicted CGM stripping condition, which motivated our modeling of the \texttt{weak-w} orbit (orange symbol; see \S \ref{subsec:orbits}).}
    \label{fig:Pram_analytical_thresh}
\end{figure}

To evaluate the condition for dwarf CGM removal, we assume that the dwarf galaxy's gravitational restoring force comes from the dark matter component alone (the gas and stellar masses are $\sim 10\%$ and $2\%$ of the dark matter mass, both are ignored here for simplicity). The gravitational potential of the dark matter halo is spherically symmetric, so the maximized acceleration is given by $a_{\rm grav,max}(R) = a_{\rm grav}(R/\cos(\theta)) \cos(\theta)$, where $\theta=45^{\circ}$, or equivalently, $a_{\rm grav,max}(R)=a_{\rm grav}(\sqrt{2} R)/\sqrt{2}$ (see illustration in Fig. 3 of \citealt{mccarthy_ram_2008}) --- the factor of $\sqrt{2}$ comes from projection along the ram pressure direction. Equation \ref{eqn:mccarthy_pressure_balance} can be rewritten as the $P_{\rm ram}$ threshold for stripping at each projected radius $R$,

\begin{equation}\label{eqn:mccarthy_pram_threshold}
\begin{split}
    P_{\rm ram,thresh}(R) & = a_{\rm grav,max}(R) \cdot \Sigma_{\rm gas}(R)\\
    & \approx \left(\frac{1}{\sqrt{2}}\right) a_{\rm grav}(\sqrt{2} R) \cdot \Sigma_{\rm gas}(R) \\
    & \approx \frac{G M_{\rm DM}(\sqrt{2} R)}{\sqrt{2} \left(\sqrt{2} R \right)^{2}} \cdot \Sigma_{\rm gas}(R),
\end{split}
\end{equation}
where the gravitational term ($GM/r^{2}$) is the enclosed dark matter acceleration at a spherical radius $r=\sqrt{2} R$, which can be expressed analytically for our dwarf galaxy (\S \ref{subsec:dwarf}); see \cite{burkert_structure_1995}. The gas surface density profile ($\Sigma_{\rm gas}(R)$) can be evaluated numerically by projecting the gas volume density at the simulation initial condition (Figure \ref{fig:dwarf_init_condition}); also see Section \ref{subsec:stripping_criteria} for detailed methodology. Figure \ref{fig:Pram_analytical_thresh} shows the predicted ram pressure threshold for our WLM-like dwarf galaxy.

The \cite{mccarthy_ram_2008} stripping criterion (based on equation \ref{eqn:mccarthy_pram_threshold}) is evaluated at all projected radii in our simulated dwarf, shown by the black solid curve in Figure \ref{fig:Pram_analytical_thresh}. The condition for complete CGM stripping is dictated by the innermost CGM, where the gravitational restoring force and gas projected density are both maximized. For our WLM-like dwarf galaxy, the resulting ram pressure threshold is $P_{\rm ram,thresh} \approx 3 \times 10^{-14} $ dyne/cm$^{2}$ at the disk-halo interface ($R=r_{\rm disk}/\sqrt{2} \approx 3.7$ kpc, right edge of the shaded disk region). This analytical CGM stripping threshold is adopted as the peak orbital ram pressure in one of our orbits (\S \ref{subsec:orbits}). We slightly enhanced the simulation value to $P_{\rm ram,peak}\approx 4 \times 10^{-14} $ dyne/cm$^{2}$ (horizontal orange line) to account for potential under-estimations, e.g., due to only including the dark matter mass in equation \ref{eqn:mccarthy_pram_threshold}. This $P_{\rm ram}$ value for the \texttt{weak-w} orbit is 10 times lower than the MW fiducial orbit in our simulations (Table \ref{tab:orbits}); it is also low compared with the typical infalling dwarfs in the Local Group \citep{putman_gas_2021}.

In Figure \ref{fig:Pram_analytical_thresh}, we also show the central dense gas stripping criterion for this dwarf galaxy (\citealt{mori_gas_2000}; see \S \ref{subsec:analytic_core_strip}) as a constant value in the core region (horizontal dashed line). This value ($P_{\rm ram} \approx 10^{-12}$ dyne/cm$^{2}$) is highly consistent between \cite{mori_gas_2000} and \cite{mccarthy_ram_2008}, both predicting that the satellite cannot be completely stripped. The radial dependence of $P_{\rm ram,thresh}$ \citep{mccarthy_ram_2008} is less applicable in the dark matter core, where the gravitational acceleration \textit{increases} with radius, and the $a_{\rm grav,max}$ approximation in equation \ref{eqn:mccarthy_pram_threshold} no longer holds. Outside of the core region, $P_{\rm ram,thresh}$ decreases with radius (solid curve in Figure \ref{fig:Pram_analytical_thresh}), which implies that despite the CGM being more massive than the gaseous disk, it is easier to strip the CGM than the disk gas based on the \cite{mccarthy_ram_2008} prediction.

To conclude, we have briefly reviewed two analytical RPS prescriptions and applied them to make predictions for our dwarf satellite stripping simulations (\S \ref{sec:methods}). Section \ref{subsec:analytic_core_strip} \citep{mori_gas_2000} predicts that the peak ram pressure in our MW fiducial orbit (Table \ref{tab:orbits}) is insufficient to remove the central dense gas in the satellite. In section \ref{subsec:analytic_cgm_strip} \citep{mccarthy_ram_2008}, we calculated the ram pressure threshold required for dwarf CGM removal (Figure \ref{fig:Pram_analytical_thresh}), which is lower than MW fiducial, and used as the peak ram pressure for one simulation orbit (\texttt{weak-w}; \S \ref{subsec:orbits}). We will further compare the analytical criteria and the simulation results in the following sections, particularly \S \ref{subsec:stripping_criteria}.

\section{Results}\label{sec:results}

This section describes our simulation results. \S \ref{subsec:gas_removal} outlines how the satellite dwarf galaxy's ISM and CGM components evolve under RPS, \S \ref{subsec:cgm_shielding} addresses whether the inclusion of the satellite CGM affects the ISM stripping rate, and \S \ref{subsec:stripping_criteria} shows the spatially resolved gas density results and compares with the analytical criterion in \S \ref{subsec:analytic_cgm_strip}.

\subsection{Gas removal and survival under RPS}\label{subsec:gas_removal}

\begin{figure*}[!htb]
    \centering
    \includegraphics[width=1.0\linewidth]{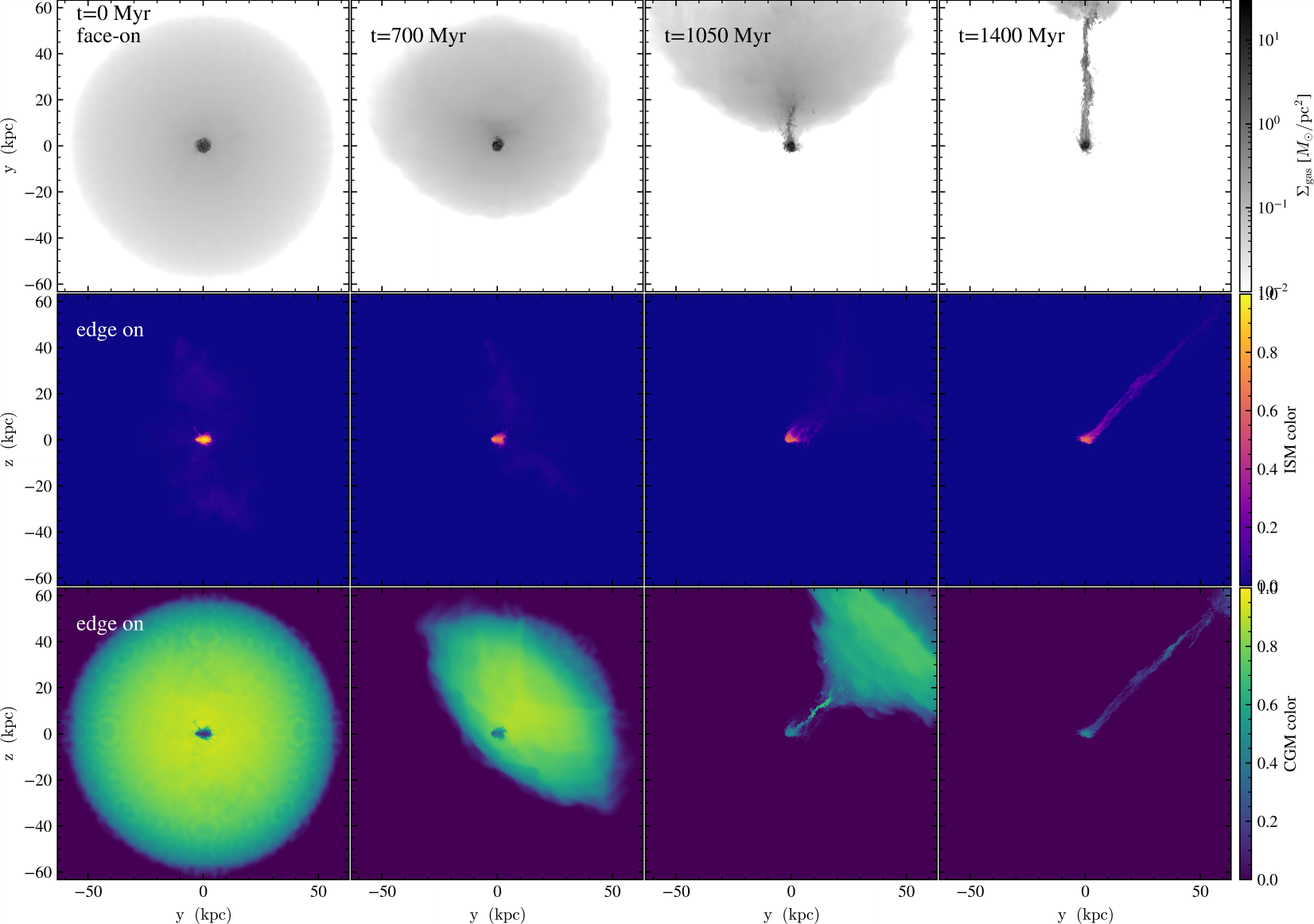}
    \caption{Gas stripping morphology of the \texttt{MW-w} simulation (Table \ref{tab:simulation_suite}). The different panels show the gas surface density ($\Sigma_{\rm gas}$, top row) and the ISM and CGM tracer fractions (middle and bottom rows; color tracers described in \S \ref{subsec:sims}) at four example snapshots (columns left to right). The $\Sigma_{\rm gas}$ panels are projected face-on (along the $z$ axis: aligned with the disk rotation axis and 45$^{\circ}$-angled with the wind), while the two color tracers are projected edge-on (along the $x$ axis) to better show global morphology under the 45$^{\circ}$ wind. Most of the CGM is quickly removed before effective ISM stripping takes place (third column), as elaborated in \S \ref{subsec:gas_removal}.}
    \label{fig:morphology}
\end{figure*}

\begin{figure}
    \centering
    \includegraphics[width=1.0\linewidth]{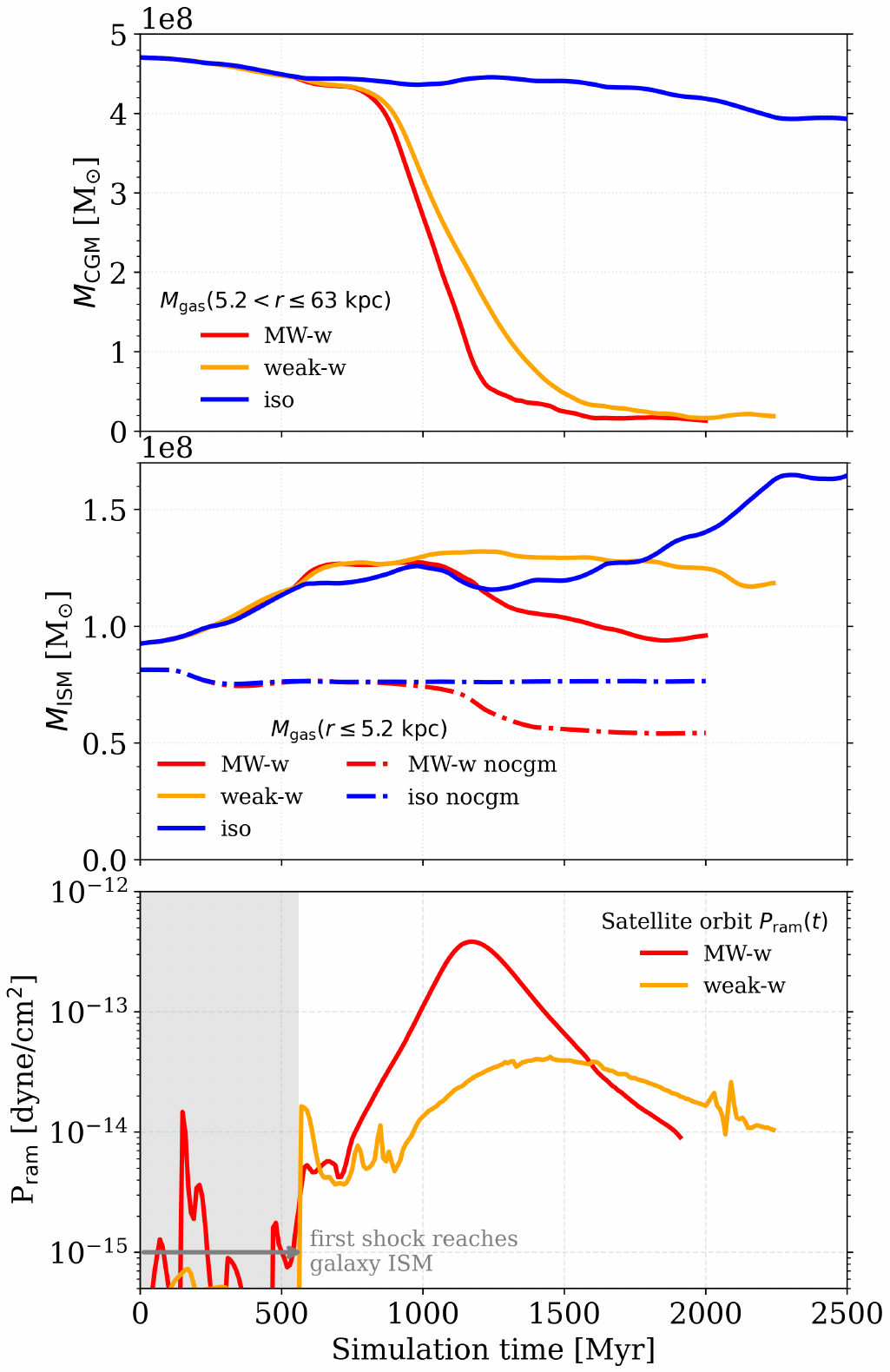}
    \caption{Dwarf satellite gas mass and orbital ram pressure time evolution. Top panel: gas mass within the CGM region, defined as $r_{\rm disk,0} < r < 1.5 r_{\rm vir}$ (the initial gas disk radius to $1.5$ times the virial radius; see \S \ref{sec:methods}). The lines show the two wind runs and \texttt{iso} control with satellite CGM. Middle panel: gas mass within the ISM region, defined as $r \leq r_{\rm disk,0}$, showing the runs with CGM as solid lines and those without CGM as dash-dotted lines in the same color. Bottom panel: ram pressure of the \texttt{MW-w} and \texttt{weak-w} orbits (host $R_{200}$ to $R_{\rm peri}$ and back to $R_{200}$; \S \ref{subsec:orbits}). The initial $\sim$550 Myr before the first shock reaches the central galaxy is shaded in gray.}
    \label{fig:global_time_evolution}
\end{figure}

We first describe the evolution of the dwarf satellite's gas content under RPS. Figure~\ref{fig:morphology} shows the gas stripping morphology of the \texttt{MW-w} run, which is then quantified in Figure \ref{fig:global_time_evolution}. 
For the runs where the satellite galaxy has a CGM (Tables \ref{table:dwarf_model} and \ref{tab:simulation_suite}), its gas mass is constituted by both ISM and CGM components, individually tracked by the fluid color tracers (\S \ref{subsec:sims}). As discussed in Section \ref{subsec:dwarf}, our galaxy evolves for 2 Gyr to allow for gradual radiative cooling in the CGM. In all the following figures, $t=0$ is when the wind enters the box immediately after that settling period. In all of our simulations, the satellite under RPS has its ISM partially removed ($<35\%$) while the CGM is almost completely removed ($>95\%$), as detailed below.


Figure \ref{fig:morphology} shows the satellite's gas surface density and its fractional components (ISM and CGM fractions) for the \texttt{MW-w} simulation (Table \ref{tab:simulation_suite}). The dense gas, i.e., the central $\sim$5 kpc and the stripped tail shown in the top row, is consistently cool ($T \lesssim 10^{4}$ K). The ISM and CGM color tracers in the middle and bottom rows show the density fractions of gas that were initially the satellite's ISM and CGM, respectively. E.g., an ISM color of $0.6$ and a CGM color of $0.3$ means the gas density along line-of-sight is $60\%$ ISM, $30\%$ CGM, and the remaining $10\%$ comes from the host wind. We have excluded wind contamination in the top row ($\Sigma_{\rm gas}$) by only showing gas with a summed ISM and CGM color greater than 0.95 (less than $5\%$ wind). The columns from left to right summarize the evolution with four characteristic time frames:

\begin{enumerate}[label=(\roman*)]
    \item Initial condition (t=0 Myr). The satellite galaxy is initially unperturbed by RPS, but its ISM and CGM interact internally. Star formation-driven feedback ejects diffuse hot ($T \sim 10^6$ K) gas into the CGM region, while some of the inner CGM cools and flows into the ISM region, leading to mixing of the color tracers.
    
    \item Onset of CGM stripping (t=700 Myr). Ram pressure-driven shock front has reached one side of the CGM, sweeping it along the 45$^{\circ}$ wind direction, $\hat{v}_{\rm wind(x,y,z)}=(0, \frac{\sqrt{2}}{2}, \frac{\sqrt{2}}{2})$. The central ISM remains largely undisturbed but at a lower ISM fraction than that in (i) because CGM inflows have continuously mixed into the ISM for 700 Myr.
    
    \item CGM mostly removed; onset of effective ISM stripping (t=1050 Myr; $\sim$150 Myr before pericenter). Most of the CGM in the inner $\sim$20 kpc radius has been stripped (except the central mixed gas), forming a stream-like tail. The ISM tracer begins to show morphological disturbances, but the extended tail gas is CGM-dominated.
    
    \item Post pericenter condition (t=1400 Myr). The low-density CGM is completely removed. The stripped tail now consists of stripped ISM and CGM, dominated by CGM at large radii and ISM closer to the dwarf.
\end{enumerate}

The physical processes in \texttt{MW-w} (Figure \ref{fig:morphology}) are representative of those in the other simulations in our suite (Table \ref{tab:simulation_suite}). For \texttt{weak-w}, the stripping morphology and time evolution closely resemble \texttt{MW-w}, despite its peak ram pressure being $\sim$10 times lower (\S \ref{sec:methods} and Figure \ref{fig:global_time_evolution} below). This agreement aligns with our expectations from our analytical calculations in Section \ref{subsec:analytic_cgm_strip}. For the \texttt{iso} control case with satellite CGM, the overall morphology is similar to the leftmost column in Figure \ref{fig:morphology} (initial condition for RPS cases), while its time evolution is dominated by the gradual mixing of the CGM into the central ISM region via inflows throughout the $\sim$2 Gyr simulation. The \texttt{MW-w} run without satellite CGM (\texttt{MW-w nocgm}) follows the ISM stripping morphology in Figure \ref{fig:morphology}'s middle row, where the onset of effective ISM stripping occurs at approximately the same time (t=1050 Myr, third column).

One unique aspect of the stripping simulations with a satellite CGM is the tail morphology; see Figure \ref{fig:morphology}'s top and bottom panels at t=1050 Myr. The wind compresses the stripped CGM gas behind the galaxy disk, leading to a dense CGM tail that is narrower than the stripped tail at later times ($t=1400$ Myr). We discuss this in more detail in Section \ref{discussion:tail}.


Figure \ref{fig:global_time_evolution} quantifies the gas removal. Consistent with observational work, for mass calculations, we distinguish between the satellite CGM and ISM (Figure \ref{fig:global_time_evolution} top and middle panels) based on purely spatial cuts, i.e., distance $r$ to the galaxy's center. E.g., stripped ISM in the tail is counted in $M_{\rm CGM}$; CGM that cools and inflows into the center is counted in $M_{\rm ISM}$. We adopt the initial disk radius $r_{\rm disk,0}=5.2$ kpc as the ISM-CGM boundary throughout this work, which closely matches the true ISM-CGM boundary in \texttt{iso} (Figure \ref{fig:siggas_compre}), but is slightly greater than the ISM radii in the RPS cases where gas is truncated. Increasing or decreasing this boundary radius does not affect overall trends.

We first discuss the satellite CGM removal quantified in the top panel of Figure \ref{fig:global_time_evolution}.  
Most notably, the satellite CGM removal is fast and effective.  In the \texttt{MW-w} and \texttt{weak-w} cases, $M_{\rm CGM}$ is rapidly lost starting at $\sim$850 Myr, with $90\%$ of $M_{\rm CGM}$ lost within $\sim$350 and 600 Myr, respectively, and the final mass loss fractions reach $>95 \%$. The true rate at which the CGM becomes unbound is even faster because our CGM spatial cut covers a large volume ($1.5 r_{\rm vir}$), and gas can remain in the volume (hence counted in $M_{\rm CGM}$) for hundreds of Myrs after it becomes unbound. For example, at t=1050 Myr, Figure \ref{fig:global_time_evolution} shows that the \texttt{MW-w} case $M_{\rm CGM}$ is $\sim 50\%$ lost, but the remaining $50\%$ resides in the unbound tail and is about to be removed (Figure \ref{fig:morphology}, third column). The final CGM mass ($\approx 2\times 10^{7}$ $M_{\odot}$) in both wind cases mostly comes from the gravitationally unbound stripped tails within $1.5 r_{\rm vir}$.

We now shift our focus to the ISM mass. The effect of RPS alone on the ISM mass evolution is shown by the \texttt{nocgm} cases (Figure \ref{fig:global_time_evolution} middle panel, dash-dotted lines): 
$M_{\rm ISM}$ remains approximately constant until $t \approx 1000$ Myr, the onset of effective ISM stripping, at which point the \texttt{MW-w nocgm} case begins to lose mass due to RPS, resulting in a $\sim 27\%$ lower final mass than the \texttt{iso nocgm} control.

In the presence of a satellite CGM, the ISM mass evolution is affected by both RPS and the ISM-CGM interactions (Figure \ref{fig:global_time_evolution} middle panel, solid lines). In all cases with a gaseous halo, the CGM adds mass to the ISM.  In \texttt{iso} (blue solid line) with our satellite CGM setup, the cool inner-CGM gas continuously inflows and replenishes the ISM, leading to the growth of $M_{\rm ISM}$. In the two stripping cases (red and orange solid lines), $M_{\rm ISM}$'s increasing trend stops at $t \approx 700$ Myr, indicating the CGM inflows are being cut off. The short-term enhanced $\Delta M_{\rm ISM} \approx 8 \times 10^{6}~M_{\odot}$ in the stripping cases relative to \texttt{iso} during $t=600-700$ Myr originates from ram pressure compressing part of the CGM gas into the ISM region (see top panel, $\Delta M_{\rm CGM} \approx - \Delta M_{\rm ISM}$ during this phase).

Subsequently, ram pressure begins to impact the ISM. $M_{\rm ISM}$ for \texttt{MW-w} begins to decrease at $t \approx 1000$ Myr, marking the effective ISM stripping phase (coincides with \texttt{MW-w nocgm}); for \texttt{weak-w}, it remains almost constant because the low $P_{\rm ram}$ is insufficient for dense gas removal. The distinct trends in \texttt{weak-w} and \texttt{iso} at $t > 1700$ Myr show that, although weak RPS causes negligible ISM removal, it affects $M_{\rm ISM}$ by removing the CGM, preventing further inflows from this gas reservoir.

The effective ISM stripping phase in \texttt{MW-w} ($1000-1400$ Myr) coincides with when the orbital ram pressure is highest ($P_{\rm ram} \gtrsim 10^{-13}$ dyne/cm$^{2}$; Figure \ref{fig:global_time_evolution} bottom panel), which corresponds to a satellite-host distance closer than 77 kpc in the \texttt{MW-w} orbit 
(\S \ref{subsec:orbits}). This phase includes the final stage of the first infall segment (approaching $R_{\rm peri}$, $P_{\rm ram}$ increases) and the initial stage of the post-pericenter segment (moving away from $R_{\rm peri}$, $P_{\rm ram}$ decreases).

\subsection{Does the satellite CGM shield its ISM against stripping?}\label{subsec:cgm_shielding}
We showed in \S \ref{subsec:gas_removal} that most of the satellite CGM mass is removed even in a low ram pressure orbit. This indicates that $M_{\rm CGM}$ is likely negligible in the 
$z \approx 0$ satellite dwarfs. 
However, if the existence of the CGM decreases the ISM stripping rate, even though it is ultimately removed, the CGM can still affect the satellite's global properties (see the Introduction). This section explores the potential effect of CGM \textit{shielding} the ISM against stripping, where we compare the ISM mass loss rates between the two \texttt{MW-w} simulations with and without satellite CGM.

\begin{figure}[!htb]
    \centering
    \includegraphics[width=1.0\linewidth]{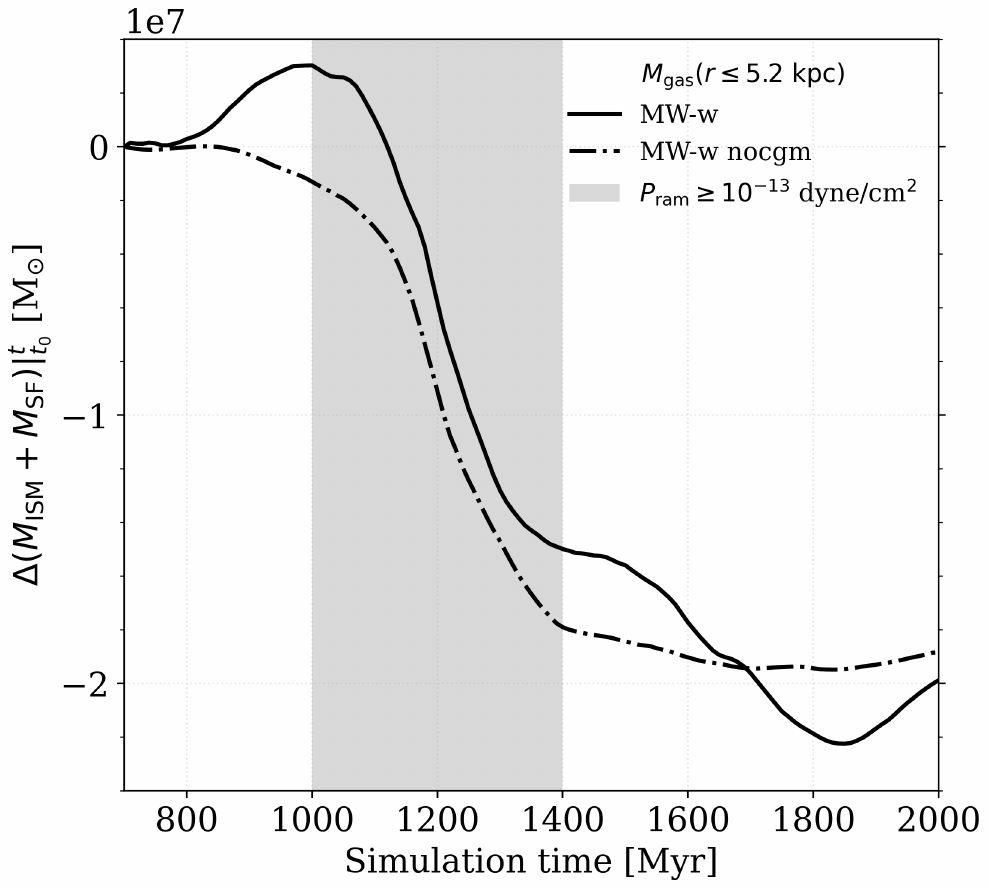}
    \caption{Dwarf satellite ISM mass loss curves under the Milky Way fiducial wind, showing $\Delta M|_{t_{0}}^{t}$: the ISM mass at time $t$ minus that at a pre-ISM stripping time ($t_{0}=700$ Myr, Figure \ref{fig:morphology}). $\Delta M$ is corrected for gas consumption by star formation; see \S \ref{subsec:cgm_shielding}. The shaded region annotates the effective ISM stripping phase (\S \ref{subsec:gas_removal}). The similar ISM mass loss rates with and without a CGM suggest that the CGM shielding effect is weak.}
    \label{fig:ism_mass_loss}
\end{figure}

Figure \ref{fig:ism_mass_loss} shows the ISM mass loss relative to an initial time frame, $t_{0}=700$ Myr, before the effective ISM stripping begins (shaded; also see \S \ref{subsec:gas_removal}). The y-axis quantity $\Delta (M_{\rm ISM} + M_{\rm SF})$ is purely kinetic, i.e., driven by mass transport alone, where the values of $M_{\rm ISM}$ in \texttt{MW-w} (solid line) and \texttt{MW-w nocgm} (dash-dotted line) are as shown in Figure \ref{fig:global_time_evolution}, and additionally corrected for gas consumption by star formation ($M_{\rm SF}$). The correction for $M_{\rm SF}$ is necessary because the SFRs for simulations with a satellite CGM are systematically higher than those without --- a direct consequence of the CGM continuously replenishing the star-forming ISM.

The ISM mass loss curves of the two cases are highly consistent in trends and values. If shielding is effective, mass loss for the case with a CGM (solid line) should be significantly lower than the one without (dashed-dotted) during the effective stripping phase, but the similar mass losses of $1.8 \times 10^{7}~M_{\odot}$ (\texttt{MW-w}) and $1.7 \times 10^{7}~M_{\odot}$ (\texttt{MW-w nocgm}) indicates that shielding is negligible. The mild mass increase in \texttt{MW-w} (solid line) before $t=1000$ Myr marks the final stage of CGM inflows, where RPS has significantly reduced the inflow rate to less than $16.7\%$ of the early steady inflow (before $t_{0}=700$ Myr, Figure \ref{fig:global_time_evolution} middle panel). To summarize, the presence of a CGM changes the ISM mass (Figure \ref{fig:global_time_evolution}) but not the ISM stripping rate (Figure \ref{fig:ism_mass_loss}).


\subsection{Radial profiles and stripping criteria}\label{subsec:stripping_criteria}

\begin{figure*}[!htb]
    \centering
    \includegraphics[width=1.0\linewidth]{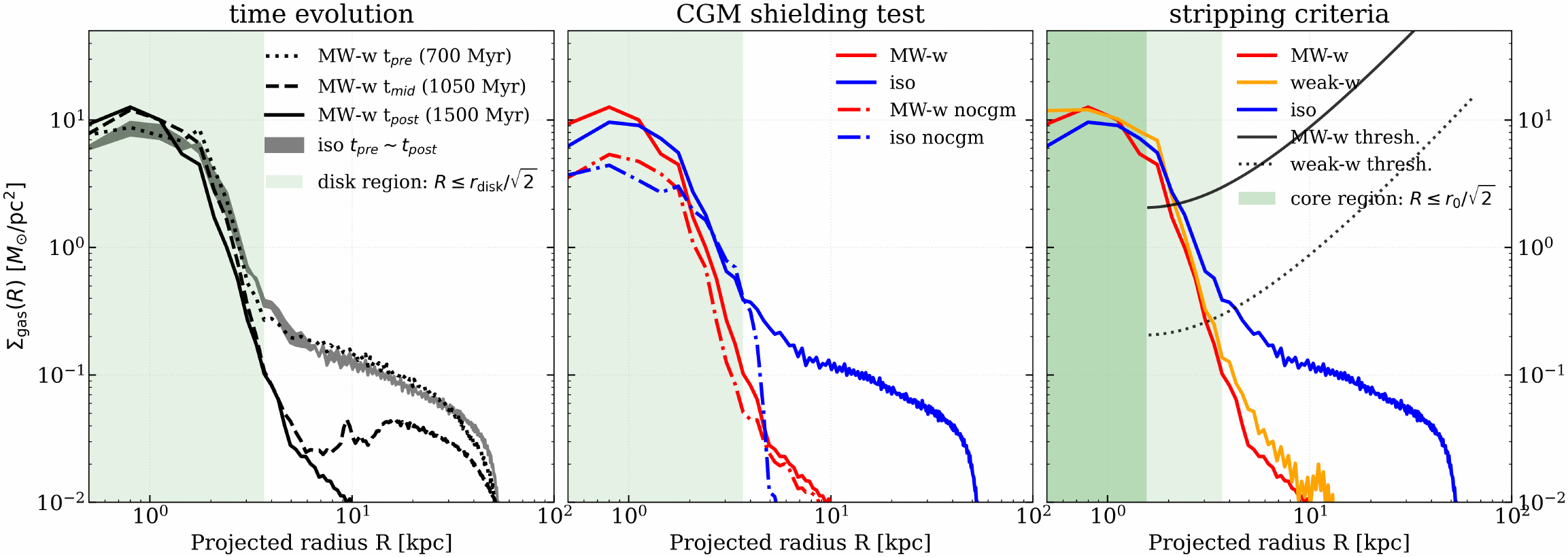}
    \caption{Azimuthally-averaged gas surface density radial profiles ($\Sigma_{\rm gas}(R)$; see \S \ref{subsec:stripping_criteria}). Left: $\Sigma_{\rm gas}(R)$ time evolution under RPS. The black curves show \texttt{MW-w} from pre- to post-pericentric times, and the filled gray curve shows \texttt{iso} over the same time range. Middle: Comparison between \texttt{MW-w} with and without the satellite CGM at $t_{\rm post}$, showing the respective \texttt{iso} control cases for reference. Lines are in the same style as Figure \ref{fig:global_time_evolution}. Right: Comparison among the three cases with CGM at $t_{\rm post}$. The \cite{mccarthy_ram_2008} stripping criteria evaluated at \texttt{MW-w} and \texttt{weak-w} peak ram pressure are shown in solid and dotted black curves, respectively, see \S \ref{subsec:stripping_criteria} for details. We shaded the projected disk region (all panels) and the dark matter core region (right panel) as in Figure \ref{fig:Pram_analytical_thresh}.}
    \label{fig:siggas_compre}
\end{figure*}

In this section, we present the effect of RPS on the gas surface density profiles. To obtain the spatially resolved profiles, we project and integrate the dwarf galaxy gas density\footnote{We include both the ISM and CGM gas of the galaxy and exclude contamination from the stripping wind using the fluid tracers introduced in Section \ref{subsec:sims} and shown in Figure \ref{fig:morphology}.} along the face-on direction (as in Figure \ref{fig:morphology} first row) and divide it into $0.32 \times 0.32$ kpc$^{2}$ square patches (our adaptive mesh resolution ranges from 80 pc in the ISM to $320-640$ pc in the CGM; \S \ref{sec:methods}). We then bin and azimuthally average the gas surface density ($\Sigma_{\rm gas}$) in the patches to obtain profiles versus the projected radius $R$. The azimuthally averaged $\Sigma_{\rm gas}(R)$ shows the mean density in the annulus, which deviates from, e.g., the stripped tail's surface density because the distribution is highly anisotropic (unlike in an unperturbed disk or CGM). Finally, the profiles are averaged over 100 Myr (10 outputs) to reduce biases from potential short-term gas displacement (e.g., due to rotation). The resulting $\Sigma_{\rm gas}(R)$ profiles are shown in Figure~\ref{fig:siggas_compre}.

The left panel of Figure \ref{fig:siggas_compre} shows the evolution of $\Sigma_{\rm gas}(R)$. Without RPS, the \texttt{iso} density profile (filled gray curve; \texttt{iso} at $t_{\rm pre}$ and $t_{\rm post}$) remains stable. Within the filled curve, there is a mild increase in the central ISM region ($\sim$3 kpc) and a mild decrease in the inner CGM ($5-15$ kpc), manifesting the gradual CGM inflows to the ISM (\S \ref{subsec:gas_removal}). Our adopted ISM-CGM boundary radius ($r_{\rm disk,0}=5.2$ kpc, shaded in all panels) agrees well with the ISM-CGM transition radius in \texttt{iso} ($4-5$ kpc).

Under RPS, the three \texttt{MW-w} time frames from pre- to post-pericentric conditions clearly show the outside-in stripping picture. From $t_{\rm pre}$ to $t_{\rm mid}$ (dotted versus dashed line), diffuse gas below $3 \times 10^{-1}~M_\odot/\rm pc^{2}$ or at projected radius $r \gtrsim 4$ kpc is rapidly removed. The remaining CGM at $t_{\rm mid}$ is already unbound and about to leave the system (Figure \ref{fig:morphology}). From $t_{\rm mid}$ to $t_{\rm post}$, as ram pressure reaches peak pericentric values (e.g., Figures \ref{fig:global_time_evolution} and \ref{fig:ism_mass_loss}), some dense gas is removed in the outer ISM region ($1.5~\rm kpc < R < r_{\rm disk,0}$; dashed versus solid line). The surface density in the central $\sim$2 kpc region never decreases as stripping proceeds; instead, it increases from $t_{\rm pre}$ to $t_{\rm mid}$ because of short-term compression of the CGM, and remains nearly constant after. 

The middle panel of Figure \ref{fig:siggas_compre} compares the runs with and without a satellite CGM at $t_{\rm post}$ --- a CGM shielding test similar to \S \ref{subsec:cgm_shielding}, here with the spatially resolved $\Sigma_{\rm gas}$ radial profiles. For the two \texttt{iso} cases (blue solid and dash-dotted curves), apart from the clear presence of CGM at $r \gtrsim 5$ kpc, $\Sigma_{\rm gas}(R)$ in \texttt{iso} is systematically higher than \texttt{iso nocgm} within the central $\sim$3 kpc because of the continuous CGM inflows over a few Gyrs replenishing the ISM (also see Figure \ref{fig:global_time_evolution} middle panel, showing the difference in $M_{\rm ISM}$). RPS removes gas from outer to inner radii (\textit{outside in}), and the stripping radius can be estimated from where the wind profile deviates from the \texttt{iso} control and decreases steeply with radius. The two \texttt{MW-w} cases (red solid and dash-dotted curves) share a remarkably consistent stripping radius of $R \approx 2$ kpc. This suggests that consistent with \S \ref{subsec:cgm_shielding}, the satellite CGM does not affect the ISM truncation radius nor shield the ISM against RPS.

The right panel of Figure \ref{fig:siggas_compre} focuses on the stripping criterion of the spherical diffuse gas (the satellite CGM), comparing our simulations with the \cite{mccarthy_ram_2008} analytic prediction (\S \ref{analytical_solutions}). The radial profiles are largely consistent between the two wind cases (red and orange solid lines); both have a stripping radius of $R_{\rm strip} \approx 2$ kpc. The \texttt{weak-w} case, with a 10 times lower peak orbital ram pressure than \texttt{MW-w}, has a slightly higher post-stripping $\Sigma_{\rm gas}$ than \texttt{MW-w} at all radii, which is also reflected in its $\sim20\%$ higher ISM mass at $t_{\rm post}=1500$ Myr (Figure \ref{fig:global_time_evolution}). Compared with \texttt{iso} (blue solid curve), the clear lack of CGM in the post-stripping wind cases confirmed the easy CGM removal, even for the \texttt{weak-w} (low ram pressure) orbit, as discussed in \S \ref{subsec:gas_removal}.

The \cite{mccarthy_ram_2008} analytic stripping thresholds (see \S \ref{analytical_solutions}, equation \ref{eqn:mccarthy_pram_threshold}) are shown in the right panel: the solid black line is evaluated at the \texttt{MW-w} peak ram pressure, and dotted black line at the \texttt{weak-w} peak ram pressure (Table \ref{tab:orbits}). Gas with surface densities below the stripping threshold is predicted to be stripped by that ram pressure. The stripping threshold adopts its radial dependence from the $1/a_{\rm grav}(R)$ term (equation \ref{eqn:mccarthy_pram_threshold}), where the gravitational acceleration $a_{\rm grav}$ is dominated by the dwarf dark matter potential
. The gas stripping density threshold (inverse of acceleration) hence increases with $R$ outside of the projected core region ($R > r_{0}/\sqrt{2} \approx 1.55$ kpc; Figure \ref{fig:siggas_compre}): it is easier to remove gas at increasing radii as the gravitational restoring force decreases with $R$.

For our simulated dwarf galaxy, the stripping radius predicted by \cite{mccarthy_ram_2008} is where the analytic stripping threshold intercepts with \texttt{iso} (blue solid line): $R=2-3$ kpc for the \texttt{MW-w} threshold and $4-5$ kpc for \texttt{weak-w}. The predicted stripping radii agree remarkably well with the \texttt{MW-w} simulation and slightly under-predict the RPS effectiveness in the \texttt{weak-w} simulation (both simulations have $R_{\rm strip} \approx 2$ kpc). The minor disagreement in \texttt{weak-w} partially arises from the steep $\Sigma_{\rm gas}$ radial dependence at $2 < R < 5$ kpc; it is challenging to estimate $R_{\rm strip}$ over these radii from the simulations. Importantly, the near complete CGM removal in our simulations is consistent with the \cite{mccarthy_ram_2008} predictions, as the predicted $\Sigma_{\rm gas}$ thresholds for survival under RPS are higher than the satellite $\Sigma_{\rm gas}$ (i.e., gas predicted to be removed) at all radii in the CGM.

\section{Discussion}\label{sec:discussion}
We can now discuss our results, beginning with general conclusions we draw based on our simulations (\S \ref{discussion:sim_vs_analytic}, \ref{discussion:inflow}, and \ref{discussion:tail}), followed by predictions for and interpretations of observations (\S \ref{discussion:massive_dwarf} and \ref{discussion:dwarf_obsn}), and concluding by noting the limitations of this work (\S \ref{discussion:limit}).

\subsection{Comparison between simulations and analytic predictions}\label{discussion:sim_vs_analytic}

We showed in \S \ref{subsec:gas_removal} that more than $90\%$ of the satellite dwarf CGM is stripped from the dark matter halo within a few hundred Myrs (even for the orbit with lower peak $P_{\rm ram}$ than typical MW satellites). The remaining $5-10\%$ of the gas mass in the halo is from the unbound gas in the tail, rather than bound CGM around the disk. Since we adopted a relatively massive satellite CGM model ($25\%$ of the baryonic mass budget; \S \ref{subsec:dwarf}), our results show that any less massive CGM or less massive dwarf galaxy will easily have its CGM stripped during its orbit.

Our simulation results are overall consistent with the instantaneous CGM stripping prediction by \cite{mccarthy_ram_2008} (\S \ref{subsec:analytic_cgm_strip}). Both orbits (Table \ref{tab:orbits}) successfully remove the dwarf CGM as predicted (Figures \ref{fig:Pram_analytical_thresh} and \ref{fig:siggas_compre}). In particular, we modeled the \texttt{weak-w} orbit's peak ram pressure to match the gravitational restoring force at the disk-halo interface. In the \texttt{weak-w} simulation, as predicted, RPS removes most of the CGM (where $P_{\rm ram} > F_{\rm grav}/dA$) and negligible ISM (where $P_{\rm ram} < F_{\rm grav}/dA$); see Figure \ref{fig:global_time_evolution}.

In the surviving ISM region, however, it is challenging to obtain accurate values of the stripping radii ($R_{\rm strip}$; see \S \ref{subsec:stripping_criteria}) because $\Sigma_{\rm gas}$ profiles decrease steeply with radius. The \cite{mccarthy_ram_2008} thresholds slightly underestimate the final $R_{\rm strip}$ in the simulations ($\sim$2 kpc; right panel of Figure \ref{fig:siggas_compre}). This suggests that the instantaneous RPS criterion for spherical diffuse gas needs modifications in the dense, turbulent, and star-forming ISM. Our result that the central ISM cannot be removed in both modeled orbits (Table \ref{tab:orbits}) agrees with predictions by \cite{mori_gas_2000} (Figure \ref{fig:Pram_analytical_thresh}).

\subsection{Implications of effective CGM removal}\label{discussion:inflow}

In this simulation suite, we have included the CGM of the satellite galaxy, which, in the absence of RPS, continuously replenishes the satellite ISM (\texttt{iso} in Figure \ref{fig:global_time_evolution}). The average CGM-to-ISM inflow rate in \texttt{iso} is $\sim$0.03 $M_{\odot}/\rm yr$, which is not intended to match the under-constrained inflow rates for dwarfs \citep{fox_gas_2017}, but a direct consequence of our massive CGM model (\S \ref{subsec:dwarf}). RPS suppresses the CGM inflows as soon as the wind reaches the ISM, halting the increasing trend of $M_{\rm ISM}$ in \texttt{MW-w} and \texttt{weak-w} ($t \gtrsim 700$ Myr; Figure \ref{fig:global_time_evolution}). The suppression of CGM inflows happens early on in the orbits, when the ram pressure is low ($P_{\rm ram} < 10^{-14}$ dyne/cm$^{2}$) and the gas stripping is yet to be effective. For our modeled dwarf satellite, RPS removes its CGM altogether and prevents further inflows.

The cessation of gas inflows onto the disk, which can lead to a gradual decrease in the satellite star formation, is referred to as \textit{starvation} \citep{larson_evolution_1980} or \textit{strangulation} \citep{balogh_h_2000} in the literature (also see, e.g., \citealt{yoon_influence_2013,peng_strangulation_2015,garling_dual_2024}). Rather than RPS being a separate quenching mechanism, our simulations confirm that RPS naturally prevents the halo gas accretion onto the disk and is connected to the starvation/strangulation mechanism \citep{cortese_dawes_2021}. Our results suggest satellite dwarf galaxies will have significantly lower CGM inflow rates and shorter quenching timescales (because of the missing gas reservoir) than field galaxies of similar masses.


\subsection{Gas tail morphology}\label{discussion:tail}

In our simulations with a satellite CGM, we find the formation of a dense, narrow tail behind the galaxy (Figure~\ref{fig:morphology}'s third column) that is formed promptly as the inner CGM is stripped away. Remarkably, this tail is composed mostly of CGM material\footnote{At least to begin with -- at later times, the CGM tail is swept away, and the tail is increasingly made of ISM stripped directly from the galaxy -- see the fourth column of Figure~\ref{fig:morphology}}. This can be seen by inspection of the ISM and CGM color fractions in the second and third rows.
The tail is not formed from stripped gas that began very close to the galaxy, as indicated by the gas tail's near-zero velocity along the wind direction. Instead, visual inspection of other outputs confirms that it is formed by radiative cooling of CGM gas that is compressed perpendicularly to the tail direction. This compression occurs because, as the CGM gas is swept away, it is done so more rapidly at larger radius (e.g., see the $t=700$ Myr edge-on view in Figure~\ref{fig:morphology}), driving a focusing effect that results in enhanced pressure in the CGM directly behind the galaxy. 

To the best of our knowledge, a tail like this, formed out of CGM (or ICM) gas, has not previously been seen in stripping simulations, which have all been of larger halos. We argue that this is directly due to the behavior of radiative cooling for the low-temperature CGM gas we are modeling. 
The CGM gas out of which the tail forms has a temperature of $T \approx 10^{5}$ K and a metallicity of $Z \approx 0.11 Z_{\odot}$. At these temperatures and metallicities, adiabatic compression (which is approximately correct in the diffuse CGM where any shocks have low Mach numbers) actually drives enhanced cooling. In particular, we find that $\partial t_{\rm cool}/\partial P$ at fixed entropy is strongly negative only in a small range of temperatures from $10^{4.7}$ K to $10^{5.2}$ K. This compression-driven cooling tail would therefore be unique to the CGM of galaxies in this mass range and would not be seen for higher mass systems.

\subsection{Predictions for the most massive dwarf satellites}\label{discussion:massive_dwarf}

We now consider the CGM stripping of the most massive dwarf satellites orbiting spiral galaxies, for example, the LMC in the MW. Being more massive than $10\%$ of the host total mass, the ram pressure threshold (equation \ref{eqn:mccarthy_pram_threshold}) of an LMC-like CGM will be an upper limit for dwarf satellites. Recent simulations \citep{lucchini_magellanic_2020,lucchini_magellanic_2021} and observational work \citep{krishnarao_observations_2022} suggest that the LMC's CGM could be the progenitor of the massive ionized component of the observed Magellanic Stream \citep{fox_kinematics_2020}.

\begin{figure}[!htb]
    \centering
    \includegraphics[width=1.0\linewidth]{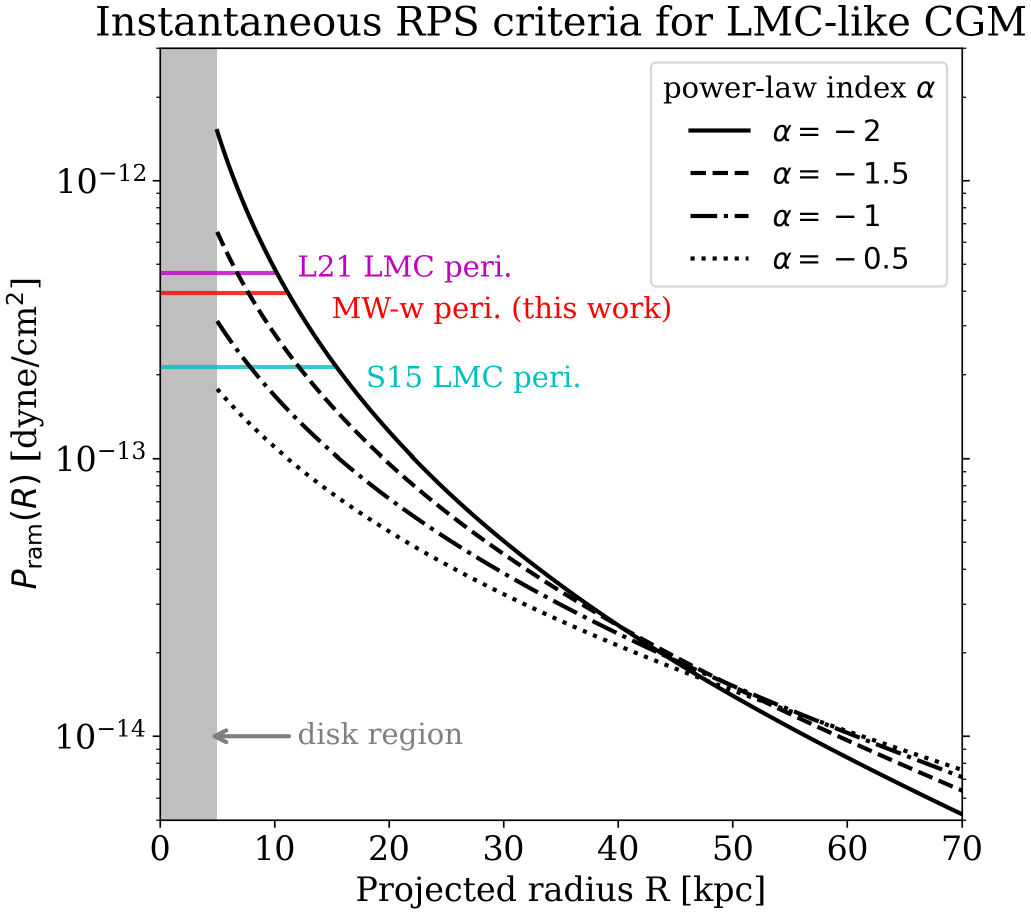}
    \caption{Instantaneous RPS criteria \citep{mccarthy_ram_2008} for an LMC-like CGM, in a similar style as Figure \ref{fig:Pram_analytical_thresh}. The line styles represent different power-law indices $\alpha$ adopted for the CGM profile (equation \ref{eqn:cgm_Menclosed}), where the total mass $M_{\rm CGM,LMC}=8.3\times 10^{9}~M_{\odot}$ follows \cite{lucchini_magellanic_2020,lucchini_magellanic_2021}. The profiles extend out to $r_{\rm vir,LMC} \approx 117~\rm kpc$ (outermost radii not shown in this figure). The horizontal lines show the $P_{\rm ram}$ values at the MW pericenter, L21: \cite{lucchini_magellanic_2021}; \texttt{MW-w}: this work (Table \ref{tab:orbits}); S15: \cite{salem_ram_2015}. In this calculation, we omit the ISM/disk region (here adopted to be $R \leq 5$ kpc; shaded).}
    \label{fig:Pram_thresh_LMC}
\end{figure}

Following the LMC parameterization of \cite{lucchini_magellanic_2021}, we assume here that the massive satellite has the NFW halo structure with $M_{200} = 1.8 \times 10^{11}~M_{\odot}$, concentration $c=9$, and initially harbors a massive CGM with $M_{\rm CGM}=8.3 \times 10^{9}~M_{\odot}$ that extends out to its virial radius ($\sim$117 kpc). The CGM stripping ram pressure can be estimated by $P_{\rm ram} > a_{\rm grav} \cdot \Sigma_{\rm gas}$ (equation \ref{eqn:mccarthy_pram_threshold}). If we adopt simplifying assumptions of the CGM density profile ($\rho_{\rm CGM}$, which integrates to $\Sigma_{\rm gas}$), e.g., following power laws as in Equation \ref{eqn:cgm_Menclosed}, we can test the CGM stripping condition for a range of characteristic density profiles, as shown in Figure \ref{fig:Pram_thresh_LMC}.

Figure \ref{fig:Pram_thresh_LMC} shows the predicted ram pressure required to remove gas at a range of projected radius, given a range of CGM models (equation \ref{eqn:cgm_Menclosed}), for the LMC-like dwarf satellite as described above. Where the ram pressure (horizontal colored lines) intersects with the analytical RPS thresholds (black curves) demonstrates the predicted CGM stripping radii. The steepest CGM mass profile ($\alpha=-2$) is the most resilient to RPS in the inner-CGM regions; it gives the upper limits of the stripping radii $R_{\rm strip}=10,11,15$ kpc, and the survived CGM mass fractions $4.6 \%, 5.5 \%, 9.3 \%$ for the pericentric ram pressure values modeled in \cite{lucchini_magellanic_2020}, \texttt{MW-w}, and \cite{salem_ram_2015}, respectively. For reference, the power-law index $\alpha$ for our modeled WLM-like satellite ranges from $-1.5$ to $-1$ (\S \ref{subsec:dwarf}), and for \cite{lucchini_magellanic_2020} $\alpha \approx -0.75$ (see their extended data figure 1). 

To summarize, with the simple approximation we have done here, we predict that most ($>90\%$) of an LMC-like massive dwarf CGM will be ram-pressure-stripped at the typical MW orbit pericenter (Figure~\ref{fig:Pram_thresh_LMC}), and there would be no detectable CGM on the LMC's leading side. If observed ionized components of the Magellanic Stream consist mainly of the stripped Magellanic Corona \citep{krishnarao_observations_2022}, the satellite is required to be on a first-infall orbit within a few hundred Myrs past pericenter.  Any previous close passage would have previously stripped the LMC's CGM.  In this first passage scenario, the LMC's \HIspace disk (ISM) should only now be affected by RPS as found by \cite{salem_ram_2015}. The \HIspace component of the Magellanic Stream is primarily linked to the 3-body tidal interaction between the LMC and SMC, and subsequently the MW \citep{besla12,lucchini_magellanic_2021}.   Given the ionized gas extends along the entire \HIspace Magellanic Stream \citep{fox14,kim_identifying_2024}, it is likely this ionized gas is not all material ram-pressure-stripped by the MW's CGM.

\subsection{Predictions for dwarf galaxy observations}\label{discussion:dwarf_obsn}

The effect of galaxy environment (e.g., isolated versus satellite) on the CGM of dwarf galaxies is under-constrained in current observations (\citealt{liang_mining_2014,bordoloi_cos-dwarfs_2014,burchett_deep_2016,johnson_extent_2017}; or see \citealt{zheng_comprehensive_2024} where the galaxy sample is selected to be isolated). We showed that the dwarf CGM, because of its lower gravitational restoring force and gas surface densities, can be $\sim$50 times more susceptible to RPS than the ISM (Figure \ref{fig:Pram_analytical_thresh}) and can be completely removed in low ram pressure orbits when the ISM is unaffected (Figure \ref{fig:global_time_evolution}). As a result, we predict that in observed gaseous/star-forming dwarf galaxies\footnote{The quenched dwarf galaxies are generally gasless; they are expected to have lost their CGM before quenching.}, the satellite population (e.g., \citealt{geha_saga_2017,mao_saga_2021,carlsten_exploration_2022,smercina_relating_2022,karunakaran_h_2022,zhu_census_2023}) is less likely to retain their CGM than isolated ``field" dwarfs at similar masses.

The predicted lack of CGM in the gaseous/star-forming dwarf satellites will affect their star formation histories (e.g., \citealt{weisz_star_2014-1,de_los_reyes_simultaneous_2022,mcquinn_jwst_2024}). Starvation will decrease the SFRs over longer timescales due to the lack of gas replenishment. However, if ram pressure affects the ISM directly, star formation could be enhanced \citep{vulcani_enhanced_2018,vulcani_gasp_2020,lee_dual_2020,roberts_lotss_2021,zhu_when_2024} or reignited \citep{wright_reignition_2019} on short timescales. We predict decreased SFRs over $\sim$Gyr timescales and note that future simulations are required to make a robust prediction of the star formation histories, which we plan to address in future work.

\subsection{Limitations}\label{discussion:limit}

We have made idealistic simplifications in our modeling choices. 
We omitted the direct modeling of magnetic fields \citep{ruszkowski_impact_2014,tonnesen_ties_2014,muller_highly_2021,roberts_radio-continuum_2023,sparre_magnetized_2024}, molecular H$_{2}$ and dust \citep{girichidis_situ_2021,farber_survival_2022,chen_survival_2023}, and cosmic rays \citep{bustard_cosmic-ray-driven_2020,farber_stress-testing_2022} in the satellite galaxy. Our setup of galaxy wind tunnel allows for high resolution in the satellite galaxy, but it omits satellite-satellite interactions or satellite-host tidal interactions, which require high-resolution cosmological simulations (e.g., \citealt{fillingham_taking_2015,simpson_quenching_2018,akins_quenching_2021,engler_satellites_2023,samuel_jolt_2023}). Finally, we assumed a smooth MW-like CGM density distribution in modeling the time-varying ram pressure, but the realistic structure of the stripping medium is expected to be more complex \citep{tonnesen_impact_2008,vandevoort_cosmological_2019,simons_figuring_2020,faucher-giguere_key_2023}. We expect our key result of the effective CGM stripping to be overall unaffected by these simplifications because the magnetic field strengths are weak, the molecular gas and dust impact is likely negligible in the CGM, and satellite-satellite interactions would only \textit{enhance} the CGM stripping in addition to RPS.

\section{Summary and Conclusions}\label{sec:summary}

In this work, we investigated the CGM stripping of an intermediate-mass dwarf galaxy ($M_{*} = 10^{7.2}~M_{\odot}$) in MW-like host environments. In the wind tunnel simulations, we varied (i) whether the dwarf satellite includes a CGM (\S \ref{subsec:dwarf}) and (ii) the ram pressure input, which represents the first-infall and post-pericentric evolution in a MW fiducial and a weak ram pressure (low eccentricity) orbit (\S \ref{subsec:orbits}); We implemented fluid tracers to independently track the dwarf satellite's ISM and CGM responses to RPS and to quantify mixing (\S \ref{subsec:sims}). For one simulation in the suite, we tailored the ram pressure input to match the analytical prediction of halo gas RPS (\S \ref{subsec:analytic_cgm_strip}) and compared the simulation results with the analytical criteria in \S \ref{subsec:stripping_criteria}. The key findings are summarized as follows.

\begin{itemize}
    \item The dwarf satellite CGM is effectively ram-pressure-stripped ($>90\%$ mass removed) within the wind crossing time of a few hundred Myrs (Section \ref{subsec:gas_removal}).
    
    \item The CGM does not affect the amount of ISM mass that is stripped or the rate at which that mass is stripped, i.e., we found no evidence for the diffuse gas shielding the central dense gas against RPS (Figure \ref{fig:ism_mass_loss}).

    \item Our simulation results are overall consistent with the \cite{mccarthy_ram_2008} analytical CGM stripping criterion, as shown from the gas surface density profiles (Figure \ref{fig:siggas_compre}).
\end{itemize}

As a result, we predict that for observed dwarf galaxies, 
\begin{itemize}
    \item Satellite dwarfs, even in low-density environments such as low eccentricity orbits around a spiral galaxy, are highly unlikely to retain their CGM at $z \approx 0$.
    
    \item Star formation for satellite dwarfs is likely in the starvation phase --- slow decrease over Gyr timescales --- because of the lack of CGM replenishment unless RPS directly impacts the ISM.
\end{itemize}

The fast and effective dwarf CGM stripping is a key finding of this work. The timescale of CGM becoming unbound is shorter than the few hundred Myr timescale where the dwarf loses most of its CGM mass (Figure \ref{fig:global_time_evolution}), because unbound gas that remains in the halo volume can still be included in a mass measurement based on spatial criteria. Given the range of ram pressure we covered (Section \ref{subsec:orbits}), the result holds for the typical infalling dwarfs in the Local Group or extragalactic spiral analogs. Indeed, orbits reaching much smaller pericentric distances would likely interact with the extended HI disk of the host, so a smooth beta profile would not be an appropriate density model.  

Importantly, we find that simulations are \textit{not} required to include a CGM in order to model the ISM stripping of dwarf galaxies.  We also find that while the \cite{mccarthy_ram_2008} analytical CGM stripping criterion is in strong agreement with our simulations, we measure stripping to smaller radii than analytically predicted, possibly hinting at the importance of feedback in enhancing gas removal from low-mass galaxies.  


\begin{acknowledgments}
GLB acknowledges support from the NSF (AST-2108470 and AST-2307419, ACCESS), a NASA TCAN award, and the Simons Foundation through the Learning the Universe Collaboration. The simulations used in this work were run and analyzed on facilities supported by the Scientific Computing Core at the Flatiron Institute, a division of the Simons Foundation.
\end{acknowledgments}

\vspace{5mm}

\software{NumPy \citep{harris_array_2020}, Astropy \citep{astropy_collaboration_astropy_2013,astropy_collaboration_astropy_2018,astropy_collaboration_astropy_2022}, yt \citep{turk_yt_2011}, and Ipython \citep{perez_ipython_2007}.}




\bibliography{dwarf_RPS,RPS,Dwarf_Gas}{}
\bibliographystyle{aasjournal}



\end{document}